\documentclass[journal=cmatex,manuscript=article,layout=twocolumn]{achemso}

\usepackage[version=3]{mhchem} % Formula subscripts using \ce{}
\usepackage{hyperref}
\usepackage{xcolor}
\usepackage{url}
\usepackage{graphicx}
\usepackage{lscape} 
\usepackage{array}
\usepackage{booktabs}
\usepackage{caption}

\newcolumntype{P}[1]{>{\raggedright\arraybackslash}p{#1}}
\author{Sakshi Agarwal}
\author{Wei Xiong}
\author{Rampi Ramprasad}
\email{rampi.ramprasad@mse.gatech.edu}
\affiliation[Georgia Institute of Technology]
{School of Materials Science and
Engineering, College of Engineering, Georgia Institute of Technology, Atlanta, Georgia 30318, United States}
\title[An \textsf{achemso} demo]
  {polyRETRO: a Language Model Approach to predict Polymerization Class and Monomer(s) for a Target Polymer}

\begin{document}

%%%%%%%%%%%%%%%%%%%%%%%%%%%%%%%%%%%%%%%%%%%%%%%%%%%%%%%%%%%%%%%%%%%%%
%% The "tocentry" environment can be used to create an entry for the
%% graphical table of contents. It is given here as some journals
%% require that it is printed as part of the abstract page. It will
%% be automatically moved as appropriate.
%%%%%%%%%%%%%%%%%%%%%%%%%%%%%%%%%%%%%%%%%%%%%%%%%%%%%%%%%%%%%%%%%%%%%
%\begin{tocentry}

%Some journals require a graphical entry for the Table of Contents.
%This should be laid out ``print ready'' so that the sizing of the
%text is correct.

%Inside the \texttt{tocentry} environment, the font used is Helvetica
%8\,pt, as required by \emph{Journal of the American Chemical
%Society}.

%The surrounding frame is 9\,cm by 3.5\,cm, which is the maximum
%permitted for  \emph{Journal of the American Chemical Society}
%graphical table of content entries. The box will not resize if the
%content is too big: instead it will overflow the edge of the box.

%This box and the associated title will always be printed on a
%separate page at the end of the document.

%\end{tocentry}

%%%%%%%%%%%%%%%%%%%%%%%%%%%%%%%%%%%%%%%%%%%%%%%%%%%%%%%%%%%%%%%%%%%%%
%% The abstract environment will automatically gobble the contents
%% if an abstract is not used by the target journal.
%%%%%%%%%%%%%%%%%%%%%%%%%%%%%%%%%%%%%%%%%%%%%%%%%%%%%%%%%%%%%%%%%%%%%
\begin{abstract}
While machine learning has transformed polymer design by enabling rapid property prediction and candidate generation, translating these designs into experimentally realizable materials remains a critical challenge. Traditionally, the synthesis of target polymers has relied heavily on expert intuition and prior experience. The lack of automated retrosynthetic tools to assist chemists, limit the rapid practical impact of data-driven polymer discovery. To expedite lab-scale validation and beyond, we present a retrosynthetic framework that leverages large language models (LLMs) to guide polymer synthesis. Our approach, which we call polyRETRO, involves two key steps: 1) predicting the most likely polymerization reaction class of a target polymer and 2) identifying the underlying chemical transformation templates and the corresponding monomers, using primarily natural-language based constructs. %We used natural language template to interpret the functional group transformation during a polymer synthesis. We demonstrate this method on condensation and addition polymerizations, achieving monomer prediction accuracies of 0.87 and 0.89, respectively. 
This LLM-driven framework enables direct retrosynthetic analysis given just the target polymer SMILES string. polyRETRO constitutes a initial step towards a scalable, interpretable, and generalizable approach to bridge the gap between computational design and experimental synthesis.
\end{abstract}
\section{Introduction}

\begin{figure*}[!ht]
	\centering
	\includegraphics[width=1\textwidth]{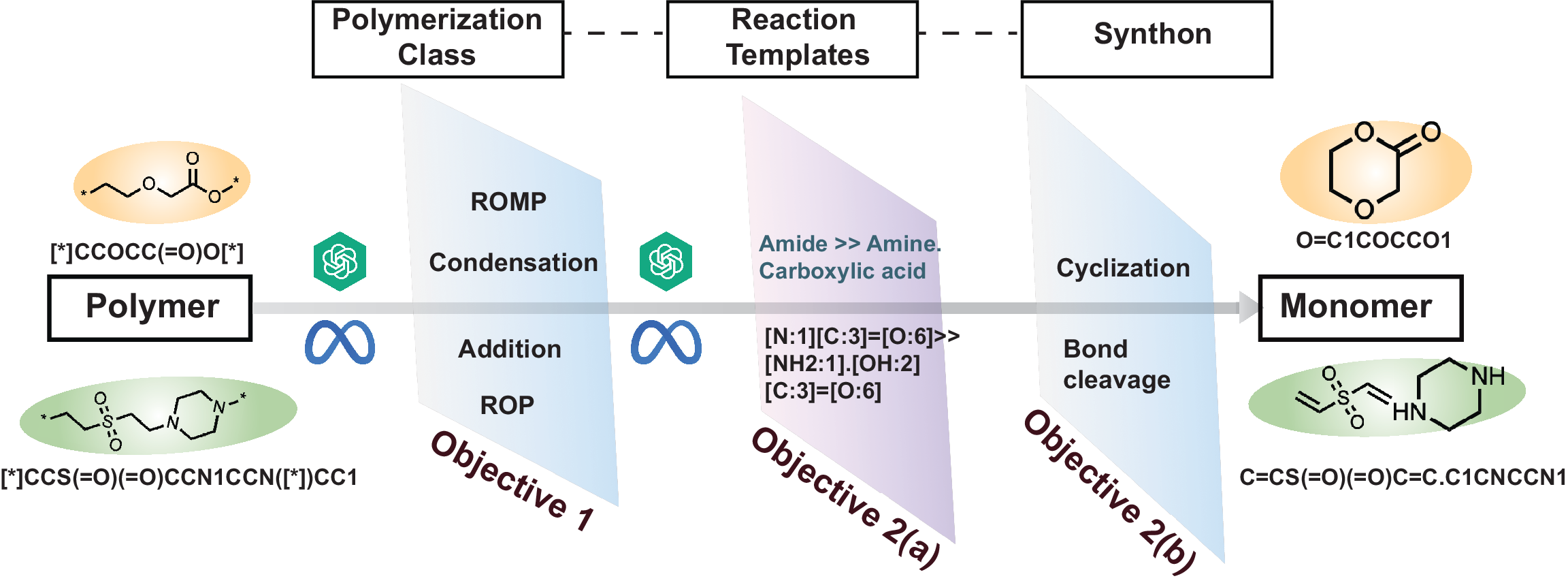}
	%\hspace{20pt}
	\caption{The retrosynthesis workflow of the polyRETRO pipeline for predicting monomers corresponding to a target polymer. Objective 1 employs a classification language model to identify the polymerization class. Objective 2(a) uses a second language model to predict the reaction template, while Objective 2(b) determines the monomers through bond cleavage or cyclization steps.}
	\label{fig1.1}
\end{figure*}
The rapid advancement of machine learning (ML) has significantly advanced the field of polymer informatics, enabling efficient prediction of polymer properties and the generation of novel candidate structures with tailored functionalities\cite{batra2021emerging,doan2020machine,kim2021polymer,kuenneth2023polybert,kern2025informatics,agarwal2025polymer}. These data-driven approaches have expanded the explored design space and reduced the reliance on trial-and-error experimentation\cite{chen2021polymer,peerless2019soft,adams2008engineering,otsuka2011polyinfo,kim2018polymer,audus2017polymer,mannodi2016rational,wu2020machine}. However, a critical bottleneck that remains is the translation of computationally proposed polymers into synthetically accessible materials\cite{tran2024design,chen2021data}. Identifying viable polymerization routes, including the appropriate selection of monomers, catalysts, solvents, and reaction conditions still heavily relies on expert domain knowledge, chemical intuition, and literature mining\cite{chen2021data}. This manual, iterative process is labor-intensive, non-systematic, and poorly suited for high-throughput autonomus workflows. Consequently, many ML-generated polymer candidates remain theoretical, with limited experimental realization. The absence of retrosynthetic analysis capabilities due to the complexity of polymer synthesis further exacerbates this gap. 

Polymer synthesis typically involves multiple stages: initial reactions between one or more monomeric units via classical organic transformations, formation of oligomers, and subsequent polymerization steps that generate long polymer chains. This hierarchical and multi-step process makes retrosynthetic mapping from a target polymer back to its monomer precursors particularly challenging. Without a reliable way to map desired polymer structures back to feasible synthetic routes, the practical utility of ML-driven polymer design is severely constrained. Bridging this disconnect between in silico design and experimental synthesis is essential for advancing the real-world impact of polymer informatics.

In contrast, retrosynthesis for small molecules has already made significant progress through both rule-based systems and end-to-end machine learning models\cite{segler2018planning,schwaller2019molecular, gao2018using,dai2019retrosynthesis, coley2017computer, sacha2021molecule, chen2023g, yao2024node}. But for polymers, retrosynthesis remains underdeveloped due to their structural complexity and the limited availability of high-quality reaction data. To address this gap, Chen et al. collected a dataset of over 10,000 polymerization reactions and developed a system that recommends potential monomers or precursors by matching the target polymer to known reaction templates via structural similarity \cite{chen2021data}. Recently, fine-tuned transformer models have also been used to predict both forward reactions and retrosynthetic paths for polymer and small molecule synthesis \cite{ferrari2024predicting, liu2024retrocaptioner, zheng2019predicting, sun2021towards, xiong2025bridging}. Rule-based tools like SMiPoly and polyVERSE have also been proposed to encode common polymerization patterns, but they do not include corresponding retrosynthetic logic\cite{ohno2023smipoly, kern2025informatics, gurnani2024ai}. Despite these advances, current methods are often limited in the types of reactions they support, lack flexible template handling, or are hard to interpret chemically, especially when it comes to polymer retrosynthesis.

To overcome these limitations for polymers, we propose a framework called polyRETRO that combines a curated set of fundamental polymerization reactions, represented through interpretable and generalizable reaction templates, with the reasoning capabilities of large pre-trained language models (LLMs). polyRETRO proceeds via two objectives based on the polymer synthesis procedure: Objective 1 fine-tunes a LLM to identify the polymerization reaction class that may be used to synthesize a given polymer; second, Objective 2 (a) involves fine-tuning of a second LLM to predict the underlying functional group transformations of the reacting monomers in the form of a reaction template. Furthermore, Objective 2(b) finally determines the monomers by locating synthons, the potential synthetic building block of the polymer through bond cleavage and cyclization. Figure 1 summarizes the process. Beginning with the SMILES representation of the polymer repeat unit, polyRETRO pipeline assigns a polymerization class, then proposes a reaction template, and finally recovers the monomers required to synthesize the target polymer. Although Objective 2(a) still leverages manually curated reaction templates, polyRetro differs significantly from traditional template matching methods, as it does not require explicit structure comparison or manual template selection. Instead, we fine-tune a LLM to directly predict a suitable reaction template, in natural language, from the SMILES string of a polymer’s repeat unit. polyRETRO retains the interpretability and chemical reasoning embedded in rule-based templates while harnessing the scalability, flexibility, and pattern-recognition strengths of LLMs. As a result, it enables a more efficient and generalizable path toward retrosynthetic analysis of computationally designed polymers, bridging the gap between in silico design and experimental feasibility.

\section{Results and Discussions}

\subsection{Objective 1 - Predicting Polymerization Class}

Objective 1 of polyRETRO focuses on predicting the class of the polymerization reaction necessary to synthesize a target polymer. Possible outcomes include condensation, addition, ring-opening polymerization (ROP), and ring-opening metathesis polymerization (ROMP).

\subsubsection{Dataset}
Our dataset of polymerization reactions contains 11,245 previously reported polymerization paths, for 10,051 homopolymers starting from 5105 unique monomers. This dataset was manually collected from various resources, including online repositories and published journal articles\cite{otsuka2011polyinfo,ma2014rationally,li2019high,wu2020flexible}. Four polymerization classes were considered, including condensation (8044 polymers), addition (2308 polymers), ring-opening (586 polymers), and ring-opening metathesis (307 polymers) as shown in Table 1. %In addition to this, the dataset also includes virtually designed polymers for condensation class ($\sim$ 20M polymers) and ROMP (28429 polymers). 
%All the polymers involved in the study are experimentally known and curated from the literature only. 
The polymers and reactant molecules are made up of 12 elements (i.e., C, H, B, O, N, S, P, Si, F, Cl, Br, and I) and a variety of polymer classes. In the present work, the role of other factors, such as solvents, catalysts, and experimental conditions, is neglected. It is worth pointing out that both the training and test datasets are composed of homopolymers and exclude copolymers, polymer blends, ladder, cross-linked, and metal-containing polymers.

\begin{table}[h!]
\centering
\caption{Summary of polymer counts per polymerization class in the dataset used for polyRETRO.}
\resizebox{\columnwidth}{!}{%
\begin{tabular}{@{}llll@{}}
\toprule
\textbf{Polymerization Class} & \textbf{Known Polymers} & \textbf{Train Polymers} & \textbf{Test Polymers} \\
\midrule
Condensation  & 8044 & 500 & 7544 \\
Addition      & 2308 & 500 & 1808             \\
Ring opening  & 586   & 500 & 86               \\
ROMP          & 307   & 250 + 250* & 57 + 28232*           \\
\bottomrule
\end{tabular}%
}
\vspace{-2pt} % tighten space if needed
\raggedright % left align the note
{\fontsize{6}{7}\selectfont * \textit{Virtual polymers}}
\end{table}

\begin{figure*}[!ht]
	\centering
	\includegraphics[width=1\textwidth]{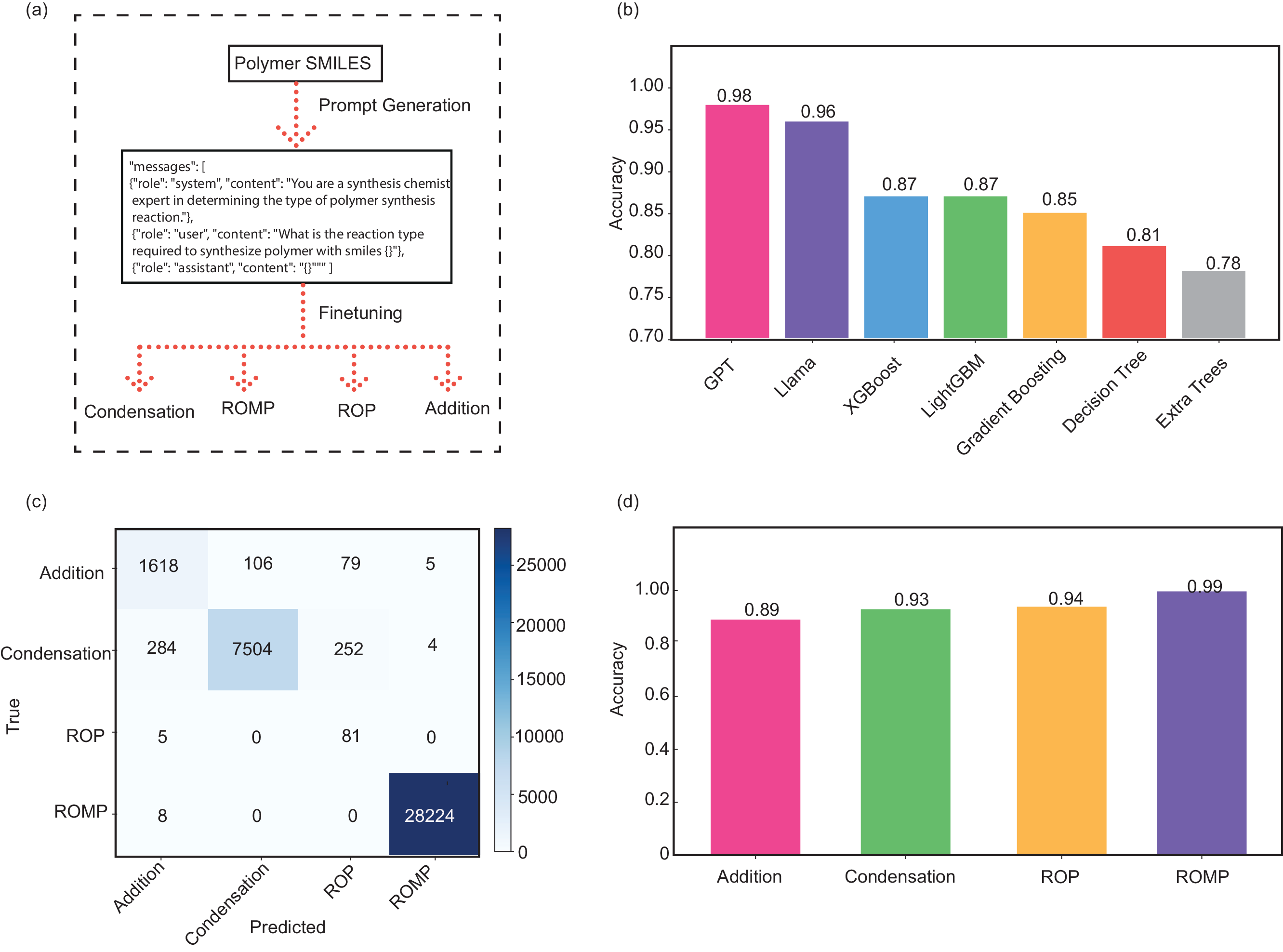}
	%\hspace{20pt}
	\caption{(a) The workflow for classifying polymerization class using LLMs, (b) The Accuracy of the LLMs and the machine learning models for the polymerization classification task, (c) The confusion matrix for the GPT fine-tuned model. and (d) The class-wise accuracy for each polymerization class for the GPT-finetuned model }
	\label{fig1.2}
\end{figure*}

The dataset used to train the model for Objective 1 of polyRETRO pipeline comprises 500 data points from each polymerization class, to ensure a balanced training set of known polymers as shown in Table 1. The remaining data points, 7544 from condensation, 1808 from addition, and 86 from ring-opening polymerization (ROP), were reserved for testing. The decision to use only 500 data points per class in the training set was made to prevent data imbalance and ensure fair model learning across all polymerization classes. In the case of ROMP, due to the limited number of curated examples, the training set includes 250 manually curated polymers and 250 virtual polymers from the total of 28,482 polymers generated in one of our previous works\cite{gurnani2024ai}. The corresponding ROMP test set consists of the remaining 57 known and 28,179 virtual ROMP polymers. 

\subsubsection{Fine-tuning}
For the classification task of predicting the polymerization class, we fine-tuned LLMs, including OpenAI’s proprietary GPT-3.5\cite{lee2020patent, brown2020language} and Meta’s open-source LLaMA-3B\cite{grattafiori2024llama}. The fine-tuning process involved training these base models on a curated dataset comprising prompt–response pairs specifically designed for the polymerization domain. Each data point was formatted as a question–answer pair consistent with the input structure expected by these LLMs. As illustrated in Figure 2 (a), the fine-tuning followed a structured message format consisting of three roles: the system role, which defines the model's expertise and sets the context for the task; the user role, which poses the question $``$What is the reaction type required to synthesize polymer with SMILES \{polymer SMILES\}?”, where the placeholder \{polymer SMILES\} is replaced with the actual SMILES (Simplified Molecular-
Input Line-Entry System)\cite{weininger1988smiles, lin2019bigsmiles} representation of the polymer; and the assistant role, which provides the correct answer. The model was trained to generate a response from a predefined set of four possible polymerization classes: “Condensation,” “Addition,” “Ring-opening,” and “ROMP.” This structured approach enabled the model to effectively learn the classification task and accurately predict the polymerization reaction class for unseen examples.

While both GPT-3.5 and LLaMA-3B were fine-tuned for the polymerization classification task using the structured prompt–response format described above, the two models differ significantly in their fine-tuning flexibility and transparency. GPT-3.5, fine-tuned via the OpenAI API, offers ease of use for training and inference but provides limited control over internal hyperparameters. As a result, GPT-3.5 functions largely as a black-box model, with fine-tuning restricted to a few configurable parameters such as the number of training epochs and the softmax inference temperature (T), which is a measure of model's imagination. Moreover, the details of the underlying fine-tuning mechanisms and model architecture remain proprietary and are not publicly disclosed. In order to optimize the number of epochs, the model was finetuned on several epochs using the 2000 train points and tested on the remaining data. Epoch 5 is found to be the most optimum. Furthermore, for inference temperature, GPT finetuned model on 2000 datapoints with epoch 5 was tested on 100 randomly selected datapoints with varying temperature. The optimum value of T was found to be  1 as shown in Figure S1. 

In contrast, fine-tuning the open-source LLaMA-3B model provided greater flexibility and control over the training process. The model was fine-tuned on in-house servers using Low-Rank Adaptation (LoRA)\cite{hu2022lora}, a parameter-efficient approach that injects trainable rank-decomposed matrices into the pre-trained weight space. This technique substantially reduced computational overhead, accelerated training time, and lowered memory usage without compromising performance. We conducted systematic hyperparameter optimization for LLaMA-3B, tuning the rank (r), scaling factor ($\alpha$), number of training epochs, and softmax temperature (T) during inference to maximize model accuracy. Detailed results of the optimization are provided in the Supporting Information. Specifically, Figure S2 presents the selected values of rank and $\alpha$, while Figure S3 highlights that 10 training epochs produced the best performance. This level of control and transparency afforded by LLaMA-3B fine-tuning was advantageous for tailoring the model to the specific demands of the task in-hand. Performance of all these models was assessed using accuracy metrics which is defined as the total number of correct prediction divided by total number of predictions.

The performance of the hyperparameter optimized fine-tuned model is shown in Figure 2(b), GPT-based fine-tuned model achieves the highest accuracy (0.98), followed by LLaMA (0.96). The fine-tuned GPT-model (Figure 2 (c)) demonstrates strong class-specific performance with particularly high precision for ROMP (true positive count: 28,224, out of 28,232) and condensation (7,504 out of 7,544), while minor misclassifications occur between addition and condensation due to structural similarities in some monomers as shown in Figure 2(c). Predictions for ROP and ROMP show high specificity with minimal cross-class confusion. The GPT model achieves per-class accuracies of 0.89 for addition, 0.93 for condensation, 0.94 for ROP, and 0.99 for ROMP, indicating consistent performance across all reaction classes and especially strong classification of ROMP-type polymers (Figure 2 (d)). The per-class accuracies for LLaMA is depicted in Figure S4.

In addition to fine-tuning LLMs, we trained traditional machine learning (ML) classifiers to predict the polymerization class using the same dataset. This workflow began by computing Polymer Genome fingerprints\cite{doan2020machine}, which capture chemically meaningful structural and electronic features of polymer repeat units. These fingerprints were used as input features for the classification models. The full ML pipeline, including data preprocessing, model training, hyperparameter tuning, cross-validation, and evaluation was implemented using scikit-learn. A thorough comparison of multiple algorithms was performed, and the top five models based on performance metrics were XGBoost, Random Forest, Gradient Boosting, Extra Trees, and Decision Tree. All models were evaluated using stratified cross-validation to ensure balanced representation across polymerization classes and to enhance the robustness of performance estimates. Details of the models and their corresponding hyperparameters are provided in the Supporting Information. 

As shown in Figure 2(b), all traditional ML models underperformed in comparison to the LLMs. Tree-based methods such as XGBoost and LightGBM achieved an accuracy of 0.87, while ensemble learners like Gradient Boosting (0.85), Decision Tree (0.81), and Extra Trees (0.78) showed even lower performance. These results highlight the superior ability of LLMs to capture complex reaction patterns and accurately classify polymerization classes. One of the reasons for the superior performance of the LLMs is their prior understanding of the various polymerization classes, which, in turn, can affect the model performance. 
\begin{figure*}[!ht]
	\centering
	\includegraphics[width=1\textwidth]{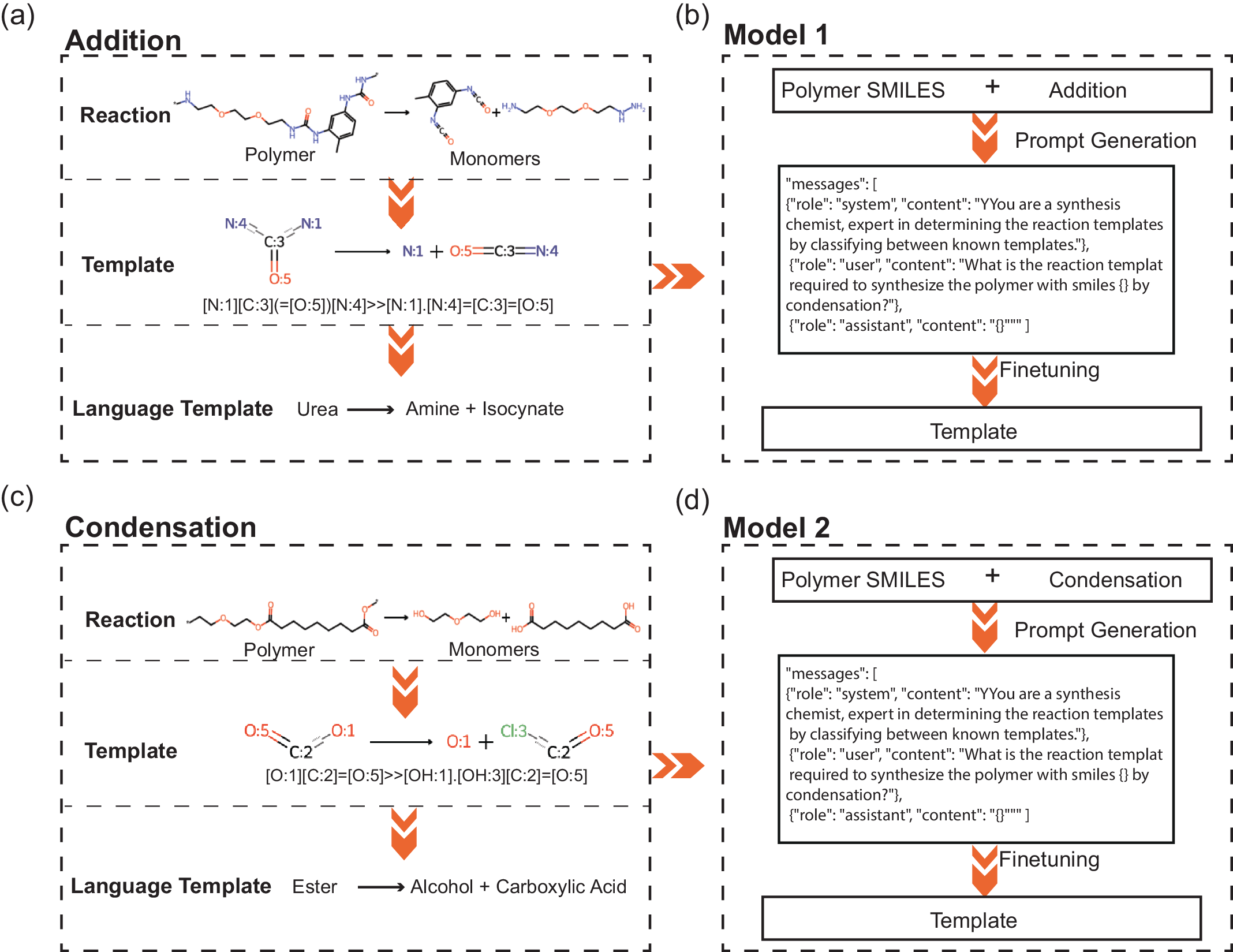}
	%\hspace{20pt}
	\caption{ (a) Template generation and (b) LLM finetuning workflow for the addition polymerization reaction. (c) Template generation and (d) LLM finetuning workflow for the condensation polymerization reaction. }
	\label{fig1.2}
\end{figure*}
\subsection{Objective 2 - Predicting Reaction Templates and Monomers}
Once the polymerization class suitable for a given polymer structure is identified, the next step of polyRETRO is to determine the corresponding monomer(s) that could have led to its synthesis. For ring-opening polymerization (ROP) and ring-opening metathesis polymerization (ROMP), this retrosynthetic prediction is straightforward, as the monomers are often cyclic compounds that can be inferred by $``$closing” the repeating unit present in the polymer backbone. In such cases, a single monomer typically undergoes ring opening to form the polymer chain. Among the 29,287 ROP and ROMP polymers evaluated in this study, the correct monomer was accurately predicted for 29,056 cases, reflecting the well-defined and predictable nature of these polymerization mechanisms.

\begin{table}[h!]
\centering
\caption{Total number of reaction templates for each polymerization class and the corresponding polymer count.}
\resizebox{\columnwidth}{!}{%
\begin{tabular}{@{}llll@{}}
\toprule
\textbf{Polymerization Class} & \textbf{SMARTS} & \textbf{language Templates} & \textbf{Total Polymers} \\
\midrule
Addition    & 129  & 42  & 2,273 \\
Condensation & 314 & 68 + 6* & 7,130 + 8000*\\
\bottomrule
\end{tabular}%
}
\vspace{-2pt} % tighten space if needed
\raggedright % left align the note
{\fontsize{6}{7}\selectfont * \textit{Virtual polymers}}
\end{table}

In contrast, addition and condensation polymerizations usually involve the coupling of two different monomeric units through more complex reaction pathways. The resulting polymer structure often does not contain explicit signatures that can directly reveal both monomer structures, making retrosynthetic inference more challenging. Accurately identifying the monomers in these cases requires understanding the specific chemical transformation involved, which cannot be inferred from the polymer alone without additional reaction context. Therefore, to enable reliable monomer prediction for addition and condensation polymers, we advanced to objective 2(a) of polyRETRO: determining the underlying reaction template. This step is crucial for disambiguating the set of possible reactants and reconstructing the polymer synthesis route accurately.

%\subsection{Objective 2}
\subsection{Objective 2(a) - Predicting Reaction Templates}
First, we focus on predicting the underlying reaction template used in the polymer’s synthesis. This step of polyRETRO serves as a critical bridge between the polymer structure and its corresponding monomeric precursors. To accomplish this, we first represent polymerization processes through generalized retrosynthetic pathways. These synthesis routes, along with their associated reaction templates, are illustrated in Figure 3 as standardized retrosynthetic schemes. Following conventional notation, the final polymer is depicted on the left side of each reaction, while the reactants, representing the monomers are shown on the right. These pathways are encoded using two complementary template formats, each designed to capture the essential chemical transformations that govern polymer formation. This framework enables the LLMs to learn the underlying reaction logic and forms the foundation for the subsequent step of monomer prediction.

The first format employs SMARTS (SMILES Arbitrary Target Specification)\cite{chen2021data} patterns, a flexible language for defining substructural transformations at the atomic level. In this representation, atoms are denoted using the format $``$[expr:n]", where expr is a valid atomic expression (e.g., element type, hybridization, aromaticity) and n is a mapping index used to track atom correspondence between reactants and products. For example, [C:2] refers to an aliphatic carbon atom labeled with the index 2. These simplified and generalizable SMARTS templates allow precise representation of the core transformation while maintaining broad applicability across diverse polymer structures. Further details on the SMARTS syntax and its application in polymer reaction modeling are provided in our previous work\cite{chen2021data}.
 
Within the polyRETRO pipeline, we also introduce natural language reaction templates to represent the same transformations in a human-readable format. This approach was motivated by the hypothesis that LLMs, which are inherently trained on natural language data, may better interpret and generalize from text-based descriptions than from symbolic representations like SMARTS. The language templates explicitly describe the functional group transformations that occur during polymerization, such as, the esterification of a carboxylic acid with an alcohol. As illustrated in Figure 3a and 3b, both addition and condensation reactions were translated into natural language templates that capture the key chemical logic. By framing reaction knowledge in a linguistically intuitive format, we aim to enhance polyRETRO's ability to reason over chemical processes and improve its performance in downstream tasks such as monomer prediction and reaction classification. 

It is important to note that the SMARTS-based representations of these reaction templates are highly detailed and structurally specific. Each SMARTS template encodes atom-level information, including hybridization, formal charge, connectivity, and atom mapping indices. As a result, the number of unique SMARTS templates corresponding to the same reaction type can be very large, since even minor structural variations across polymers lead to different SMARTS patterns. This granularity, while chemically precise, poses a challenge for generalization and model training due to the large template space. This was also one of the reasons for translating the SMARTS templates into natural language templates that abstract away atom-level details and instead focus on the functional group transformations central to the reaction mechanism. This conversion significantly reduced the number of unique templates by capturing reaction logic in a more generalizable and interpretable format. For example, multiple SMARTS templates representing variations of an amide formation reaction could be unified under a single language template such as Amide $>>$ Amine.Carboxylic acid. This simplification not only reduces the complexity of the classification task but also aligns better with the capabilities of LLMs. Therefore, the use of generalized language templates plays a crucial role in improving model interpretability, reducing data sparsity, and facilitating more robust learning.

\subsubsection{Dataset}
As summarized in Table 2, a total of 42 unique natural language templates were developed for addition reactions, covering 2,273 polymer instances, while 68 templates were created for condensation reactions, corresponding to 7,130 manually curated known polymers and 6 templates for click chemistries, corresponding to a total of 8000 virtual polymers selected for this study. The virtual polymers are added to the dataset in order to add more polymer classes and differet chemistries to the dataset. The addition and condensation reaction templates capture the underlying functional group transformations that drive polymer formation and are used to map the polymer structure to its likely synthetic route. It is important to note that the number of SMARTS templates corresponding to the language templates is significantly high as shown in Table 2. 
\begin{table*}[h!]
\centering
\caption{Total number of different polymer types and their corresponding language templates for addition polymerization.}
\resizebox{\textwidth}{!}{%
\begin{tabular}{@{}lllll@{}}
\toprule
\textbf{Polymer Type} & \textbf{Templates Count} &  \textbf{Polymer Count}& \textbf{Train polymer count} & \textbf{Test polymer count}\\
\midrule
Alkane             & 2 & 1,315 & 1184 & 131 \\
Alkene             & 5 & 209 & 188 & 21  \\
Urea               & 2 & 144 & 130 & 14  \\
Cyclicalkane       & 3 & 106 & 95 & 11  \\
Urethane           & 1 & 103 & 93 & 10   \\
Amine              & 2 & 64  & 58 & 6  \\
Alcohol            & 4 & 54  & 49 & 5  \\
Carboxylic acid    & 1 & 31  & 28 & 3   \\
Thiol              & 2 & 26  & 23 & 3  \\
Ether              & 1 & 20  & 18 & 2  \\
Aldehyde           & 4 & 20  & 18 & 2  \\
Sulfone            & 2 & 18  & 16 & 2  \\
Amine(Alcohol)     & 1 & 18  & 16 & 2  \\
Amine(Thiol)       & 1 & 14  & 13 & 1  \\
Ester              & 4 & 12 & 11 & 1  \\
Silane             & 1 & 9  & 8 & 1   \\
Silylether         & 2 & 6  & 5 & 1   \\
Nitrile            & 1 & 6  & 5 & 1   \\
Amide(Ester)       & 1 & 4  & 3 & 1   \\
Imine              & 1 & 3  & 2 & 1   \\
Disulfide          & 1 & 1  & 1 & -   \\
\bottomrule
\textbf{Total}     & \textbf{42} & \textbf{2308} & \textbf{2077} & \textbf{231}\\
\bottomrule
\end{tabular}%
}
\end{table*}

\begin{table*}[h!]
\centering
\caption{Total number of different polymer types and their corresponding language templates for condensation polymerization.}
\resizebox{\textwidth}{!}{%
\begin{tabular}{lllll}
\toprule
\textbf{Polymer Type} & \textbf{Templates Count} & \textbf{Polymer Count} & \textbf{Train Polymer Count} & \textbf{Test Polymer Count} \\
\midrule
Amide & 2 & 2973 & 2676 & 297 \\
Ester & 4 & 2263 & 2037 & 226  \\
Thioether & 2 & 16 & 15 + 500* & 1 + 500* \\
Urethane & 2 & 14 & 13 +500* & 1 + 500* \\
Ethersulfone & 1 & 1000* & 500* & 500* \\
Furan Maleimide & 1 & 1000* & 500* & 500* \\
Imide & 1 & 1000* & 500* & 500* \\
Oxime & 1 & 1000* & 500* & 500* \\
Triazole & 1 & 1000* & 500* & 500* \\
Urea & 1 & 1000* & 500* & 500* \\
Ether & 3 & 569 & 512  & 57 \\
Amine & 6 & 324 & 288 & 36 \\
Cyclicalkane & 2 & 255 & 225 & 30 \\
Imine & 4 & 122 & 109 & 13 \\
Alkene & 2 & 33 & 29 & 4 \\
Alkane & 3 & 46 & 41 & 5\\
Thiol & 2 & 48 & 43 & 5 \\
Ether(Amine) & 2 & 47 & 42 & 5 \\
Disilane & 1 & 35 & 32 & 3 \\
Alcohol & 2 & 32 & 28 & 4 \\
Aldehyde & 1 & 21 & 19 & 2 \\
Sulfonic acid & 1 & 30 & 27 & 3 \\
Silane & 1 & 13 & 12 & 1 \\
Alkyne & 1 & 18 & 16 & 2 \\
Siloxane & 1 & 15 & 13 & 2 \\
Silylether & 2 & 17 & 7 & 10 \\
Thioamine & 1 & 10 & 9 & 1 \\
Amine(Thiol) & 1 & 9 & 8 & 1\\
Disulfide & 2 & 9 & 8 & 1 \\
Thiourea & 1 & 5 & 4 & 1 \\
Sulphonamide & 1 & 5 & 4 & 1 \\
Azo & 1 & 5 & 4 & 1 \\
Thioamide & 1 & 4 & 3 & 1 \\
Amine(Thioether) & 1 & 1 &  1 & -- \\
Thiocarbamoyl & 1 & 3 & 2 & 1  \\
Ester(Amine) & 1 & 2 & 1 & 1 \\
Amide(Ester) & 1 & 2 & 1 & 1 \\
Cyanide & 1 & 2 & 1 & 1 \\
Carboxylic acid & 1 & 1 & 1& -- \\
Sulfamide & 1 & 1 & 1 &-- \\
Thioester & 1 & 1 & 1 & -- \\
Thionoester & 1 & 1 & 1 & -- \\
\bottomrule
\textbf{Total}     & \textbf{74} & \textbf{8044} &\textbf{7240 + 4000*} &\textbf{804 + 4000*}  \\
\bottomrule
\end{tabular}%
}
\vspace{-2pt} % tighten space if needed
\raggedright % left align the note
{\fontsize{6}{7}\selectfont * \textit{Virtual polymers}}
\end{table*}

A further breakdown of these language templates by polymer type is provided in Tables 3 and 4 for addition and condensation reactions, respectively. For example, the majority of addition polymers are alkanes, which are associated with two distinct reaction templates: Alkane $>>$ Alkane.Alkane and Alkane $>>$ Alkene.Alkene. These templates represent different retrosynthetic possibilities for forming alkane-based polymers, depending on the specific reactant structures involved. The most prevalent class in this category is polyamides, which are predominantly derived from two common reaction templates: Amide $>>$ Amine.Carboxylic acid and Amide $>>$ Amine.Halide. These templates reflect the two major synthetic routes for amide bond formation via carboxylic acid activation or halide substitution. A detailed description of each template for all the polymer class is provided in Table S1.

\subsubsection{Fine-tuning}

To train the models for reaction template prediction, we fine-tuned the LLMs using these two types of structured templates as target outputs.  This task builds upon the earlier classification of polymerization classes (objective 1) and represents a more fine-grained objective that enables the reconstruction of complete polymer synthesis routes. To fine-tune the LLM for reaction template prediction, the polymer–template dataset was reformatted into question–answer pairs, as shown in Figure 3. Each question was phrased as- $``$What is the reaction template required to synthesize the polymer with SMILES $<$Polymer SMILES$>$?”, where  $<$Polymer SMILES$>$ represents the specific polymer structure. The corresponding reaction template was provided as the assistant’s response. During fine-tuning, the model was trained to associate these polymer SMILES inputs with the correct template from the curated set.

As illustrated in Figure 3, we trained separate models for the addition and condensation polymerization classes (Model 1 and Model 2). To initially assess the model's ability to understand both types of template formats (SMARTS and natural language), we fine-tuned GPT 3.5 using a subset of 1,000 data points for each template format, resulting in four independent models corresponding to the two polymerization classes and the two template formats. Model performance was evaluated using accuracy, defined as the number of correctly predicted templates divided by the total number of data points. As shown in Figure S6, GPT 3.5 trained on natural language templates achieved significantly higher accuracy for both addition and condensation compared to the models trained on SMARTS templates. These results suggest that LLMs interpret human-readable chemical descriptions more effectively than symbolic SMARTS patterns. Based on this insight, we proceeded to develop the final predictive models by finetuning GPT-3.5 and  LLaMA 3B with the complete dataset of addition and condensation using natural-language templates.

\begin{figure*}[!ht]
	\centering
	\includegraphics[width=1\textwidth]{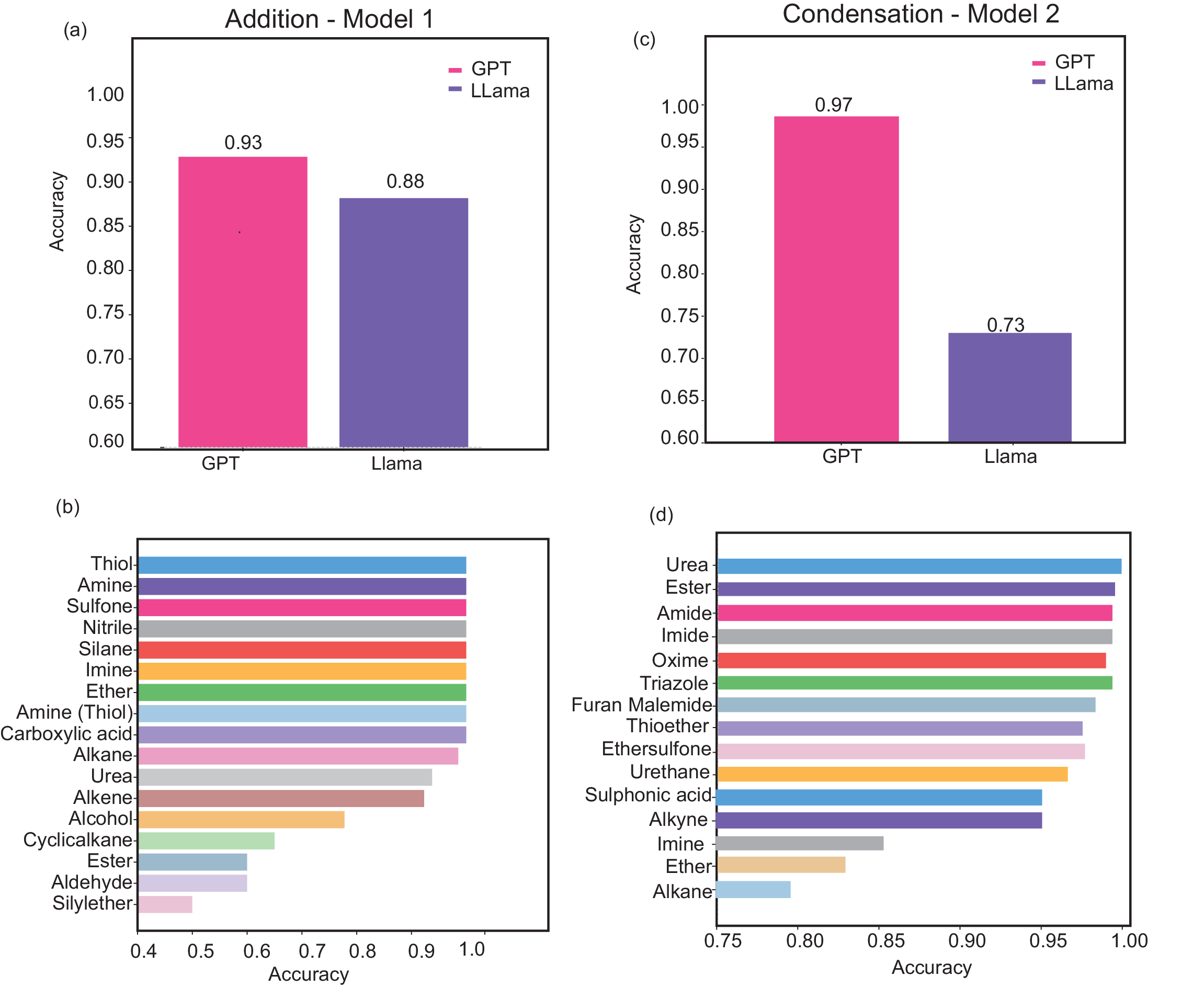}
	%\hspace{20pt}
	\caption{(a) Template classification accuracy for addition polymerization by GPT and LLaMa. (b) Accuracy for template prediction per polymer class of addition polymers using GPT.}
	\label{fig1.4}
\end{figure*}
\subsubsection{Addition Templates - Model 1 }
For the addition polymerization model (Model 1), the complete addition dataset was divided using a 90 to 10 split, where the training set included 2,002 unique polymers and the test set comprised 271 unique polymers with their corresponding language templates. The train and test set contains various polymer types, however, there is an imbalance between the different polymer types as shown in Table 3. This split was used to fine-tune both GPT 3.5 and LLaMA 3B models. Hyperparameter optimization for the GPT and LLaMA 3B model is detailed in Figure S7 and S8. Inference temperature of 0.5 was found to be the most optimum for GPT-3.5. The best-performing configuration for LLaMA finetuning was found to be a rank of 32, $\alpha$ of 64, and 20 training epochs. 

Figure 4a presents the performance of the optimized fine-tuned models, demonstrating that both LLMs are capable of accurately predicting addition reaction templates for previously unseen polymers. The GPT model achieved a high overall accuracy of 0.93, indicating strong predictive capability and generalization. LLaMA also performed well, though slightly below GPT, confirming its potential as a viable alternative. These results suggest that LLMs, particularly GPT, are highly effective at capturing the underlying reaction logic from polymer structures. Although the overall accuracy of LLaMA is lower than that of GPT, it offers several important advantages. Figure 4b provides a class-wise breakdown of accuracy for the GPT model across different polymer types. The results indicate consistently high accuracy for the major polymer classes, which have strong representation of each corresponding reaction template in the training dataset. However, for underrepresented polymer types with limited data per reaction template, the model performance decreases slightly, reflecting the influence of insufficient training examples on predictive accuracy.
For instance, silylether constitutes only 0.25 \% of the training dataset and is associated with two distinct reaction templates, while aldehyde represents about 0.9 \% of the dataset with four different reaction templates. Consequently, these sparsely represented polymer types, each associated with multiple reaction pathways but limited data coverage, exhibit poorer predictive performance compared to polymer classes with broader and more balanced representation in the dataset.
%\begin{figure}[!ht]
	%\centering
	%\includegraphics[width=1\columnwidth]{./images/Figure_5_v2.eps}
	%\hspace{20pt}
	%\caption{(a) Template classification accuracy  for condensation polymerization. (c) Accuracy for template prediction per class of  condensation polymers. }
	%\label{fig1.4}
%\end{figure}
\subsubsection{Condensation Templates - Model 2 }
Similarly, for the condensation polymerization model (Model 2), we fine-tuned LLMs using the condensation polymer dataset paired with the corresponding natural language reaction templates. The dataset comprised two components: a curated experimental known polymer set of approximately 7,000 condensation polymers corresponding to 68 reaction templates, and a second set of 8000 virtual polymers generated in-house using common condensation templates and click chemistry strategies (Table S1). The distribution of polymer types and their associated templates across the experimental and virtual datasets is summarized in Table 4. For fine-tuning, the experimentally known polymer dataset was divided into a 90/10 train–test split, ensuring that all unique reaction templates were represented within the training set across various polymer classes. Furthermore, the same 90 \% experimental training set was augmented with 500 additional virtual data points from each of the eight virtual polymer classes, incorporated into both the training and test subsets (Table 4).
This augmentation step expanded the chemical and structural diversity of the training corpus, enabling the model to generalize more effectively across novel polymer types and reaction chemistries, thereby improving the robustness of the condensation polymerization predictions.
\begin{figure*}[!ht]
	\centering
	\includegraphics[width=1\textwidth]{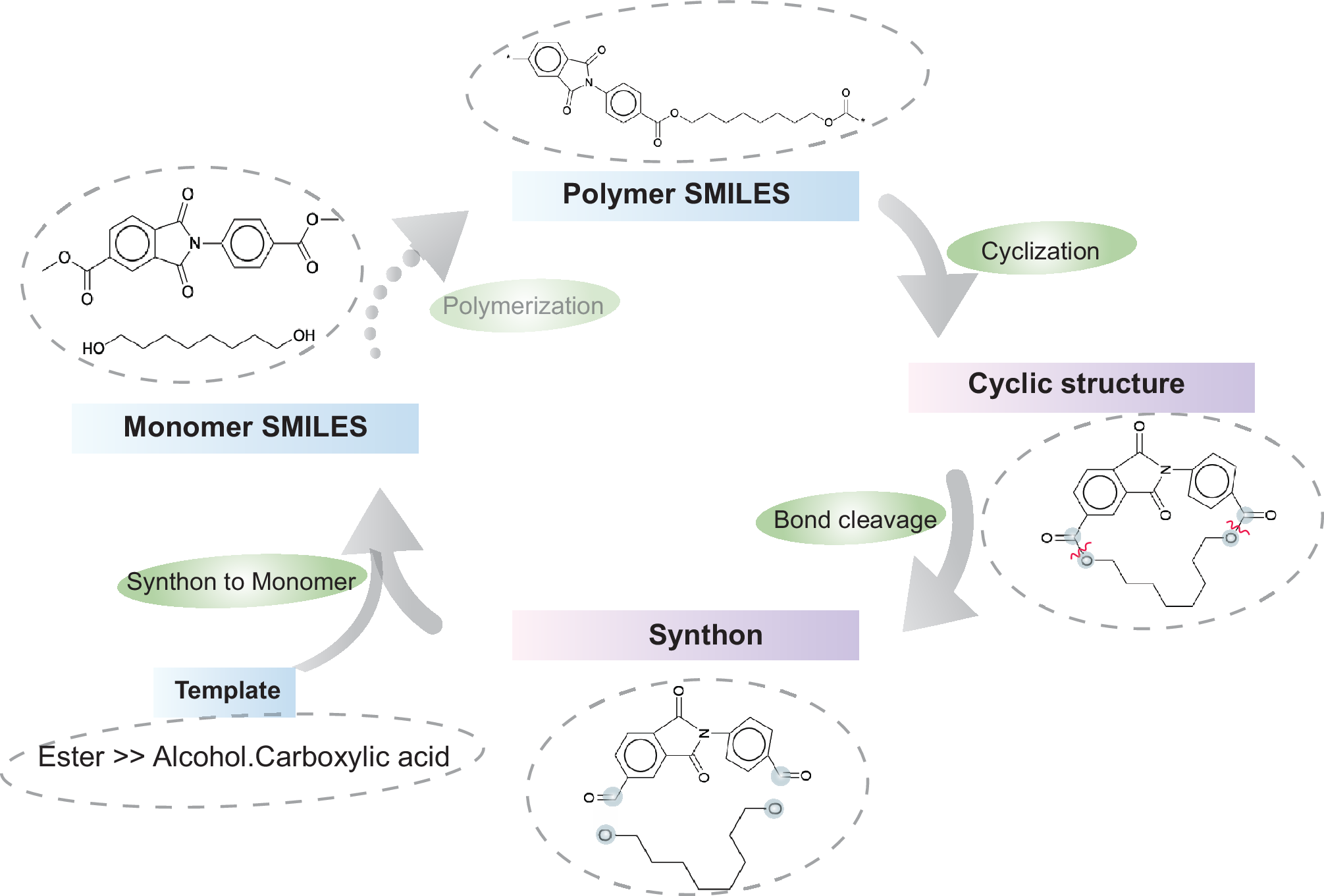}
	%\hspace{20pt}
	\caption{The workflow for Objective 2(b) of the polyRETRO pipeline illustrates the mapping of polymer SMILES to their corresponding monomer structures.}
	\label{fig1.4}
\end{figure*}

Both GPT 3.5 and LLaMA 3B were fine-tuned using these two datasets. The comparative performance of the models is shown in Figure 4c, which clearly indicates that GPT outperforms LLaMA. GPT has an accuracy of 0.97, on the other hand, LLaMA has lower accuracy of 0.73. Class-wise performance for the GPT models is shown in Figures 4d. GPT achieves high accuracy in identifying the correct reaction templates. The model performs exceptionally well for the classes having balanced 500 data points per class in both training and testing. This balanced representation ensures that the fine-tuned models are able to generalize effectively and accurately predict the reaction templates for the dominant classes of condensation polymers. Furthermore, for other classes like ester, amide, ethersulfone, sulphonic acid, alkyne, ether, etc, the model is able to effectively select the reaction templates.   
\begin{table}[h!]
\centering
\caption{The overall accuracy of all the polymer retrosynthetic steps for the best models in the polyRETRO pipeline.}
\resizebox{\columnwidth}{!}{%
\begin{tabular}{@{}llll@{}}
\toprule
\textbf{Objective} & \textbf{Prediction Task} & \textbf{Polymerization class} & \textbf{Accuracy} \\
\midrule
Objective 1    & Polymerization class   & -  & 0.98 \\
Objective 2 & Reaction template  & Addition & 0.93\\
Objective 2 & Reaction template  & Condensation & 0.97\\
Objective 2 & Monomer  & Addition & 0.89\\
Objective 2 & Monomer (Expt. $+$ virtual)  & Condensation & 0.87\\
\bottomrule
\end{tabular}%
}
\end{table}

\subsection{Objective 2(b) - Monomer Prediction}
Using the fine-tuned models (Model 1 and Model 2), we are now able to determine the reaction template corresponding to a given polymer for both addition and condensation, which is a critical step toward predicting the monomers required for its synthesis. The final step in polyRETRO involves linking the polymer SMILES and the predicted reaction template to the actual monomer structures. This process is illustrated in Figure 5. Starting from the polymer SMILES, as an initial preprocessing step, we cyclically connect the head and tail atoms of the repeat unit to simulate the polymer's continuous backbone. This ensures that all bond environments are treated symmetrically, and enables uniform cleavage logic across different polymer types. Based on the predicted template, we identify the specific bond(s) that need to be cleaved to derive the corresponding synthons. A synthon is an idealized fragment of a target molecule that represents a potential synthetic building block\cite{corey1969computer}. These fragments will have open ends, for instance, after breaking a C–O bond in an ether, we get a carbon atom and an oxygen atom, each with a free valence (C*, O*). The atomic environment around the break points is preserved, so we can attach the right end group. The reaction logic encoded in the template is then reapplied to reconstruct the full monomer structures from these fragments, i.e., synthons. For example, in the case of a polyester, a cyclic structure is first generated from the polymer SMILES. The ester linkage is then identified as the bond to be broken, as dictated by the corresponding Ester $>>$ Alcohol.Carboxylic acid template. Upon cleavage, the resulting fragments, an alcohol and a carboxylic acid functional group are interpreted as the monomers that would undergo condensation to form the original polymer. 

Final evaluation showed that the overall retrosynthesis accuracy of polyRETRO for the condensation polymerization dataset was 0.87. Here, the accuracy is defined as the correctly predicted monomers divided by the total number of datapoints. Furthermore, the polyRETRO pipeline achieved a high retrosynthesis accuracy of 0.89 on the addition polymerization dataset. The overall accuracy for each step is given in Table 5. The slight reduction in accuracy compared to template prediction arises from the strict criteria of an exact match between the predicted and ground truth monomers. In several cases, the predicted monomers differed only in their terminal groups. Therefore, such predictions should not be regarded as entirely incorrect. The accuracy does not capture these slightly different but acceptable predictions. Some examples of the predicted monomers for both addition and condensation is provided in table 7-12. The overall monomer prediction as shown in table 7-12 is quite robust, suggesting the effictiveness of the template-guided polyRETRO approach. 
%that this template-guided retrosynthetic is robust and effective for generalizing LLM-based retrosynthesis across the polymer domain. 

\subsection{Conclusion}
In conclusion, we introduce an LLM-based framework called polyRETRO, for predicting the monomers required to synthesize a target polymer. polyRETRO first determines the polymerization class (condensation, addition, ROP, or ROMP) of a target polymer using a fine tuned GPT-3.5 model, achieving an accuracy of 0.98. In the next step, for the ROP and ROMP polymers, the monomer is  directly inferred through ring-closure. In contrast, for the addition and condensation, the LLMs infers the underlying functional group transformations through natural language reaction templates that provide an interpretable description of the underlying reaction chemistry. GPT-3.5 gave the most reliable template predictions for both addition and condensation polymerizations. These natural-language reaction templates are finally mapped to their corresponding monomers, yielding a complete monomer assignment for the target polymer. The polyRETRO pipeline serves as an initial step towards a scalable, accurate, and interpretable approach to monomer prediction that could accelerate polymer design the closure of the gap between in silico polymer discovery and experimental realization.

\begin{acknowledgement}
This work was supported by the Office of Naval Research through grants N00014-19-1-2103 and N00014-20-1-2175. %Pranav Shetty was partially
%funded by a fellowship by JPMorgan Chase $\&$ Co. that helped to support this
%research. Any views or opinions expressed herein are solely those of the
%authors listed, and may differ from the views and opinions expressed by
%JPMorgan Chase $\&$ Co. or its af liates.

\end{acknowledgement}

%%%%%%%%%%%%%%%%%%%%%%%%%%%%%%%%%%%%%%%%%%%%%%%%%%%%%%%%%%%%%%%%%%%%%
%% The "Acknowledgement" section can be given in all manuscript
%% classes.  This should be given within the "acknowledgement"
%% environment, which will make the correct section or running title.
%%%%%%%%%%%%%%%%%%%%%%%%%%%%%%%%%%%%%%%%%%%%%%%%%%%%%%%%%%%%%%%%%%%%%

%%%%%%%%%%%%%%%%%%%%%%%%%%%%%%%%%%%%%%%%%%%%%%%%%%%%%%%%%%%%%%%%%%%%%
%% The same is true for Supporting Information, which should use the
%% suppinfo environment.
%%%%%%%%%%%%%%%%%%%%%%%%%%%%%%%%%%%%%%%%%%%%%%%%%%%%%%%%%%%%%%%%%%%%%
\begin{suppinfo}

Table containing detailed description of all the templates present and their corresponding polymer class. Hyperparameter optimization of GPT and LLaMa for the objective 1. Confusion matrices corresponding to the fine-tuned LLaMa models. Details and the class-wise accuracy of XGBoost. Hyperparameter optimization of GPT and LLaMa for the objective 2(a).

\end{suppinfo}

\begin{landscape}
\begin{table}[h!]
\centering
\renewcommand{\arraystretch}{1.8}
\begin{tabular}{|P{9.5cm}|P{6.5cm}|P{6cm}|}
\hline
\textbf{Polymer } & \textbf{Language Template} & \textbf{Monomers} \\
\hline
\parbox{9.5cm}{
\small\textbf{SMILES:}\\
\texttt{[\textasteriskcentered]Nc1ccccc1CCc1ccc(NC(=O)NCCCCCCNC([\textasteriskcentered])=O)cc1}\\[1ex]
\includegraphics[width=8.5cm]{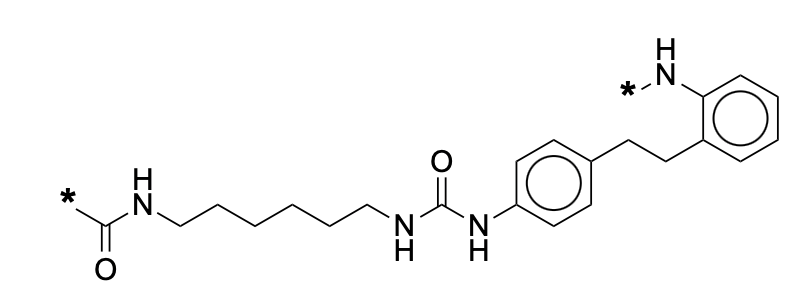}
} &
\parbox{6.5cm}{
\textbf{Addition}\\
Urea $\Rightarrow$ Amine.Isocyanate
} &
\parbox{6cm}{
\includegraphics[width=5cm]{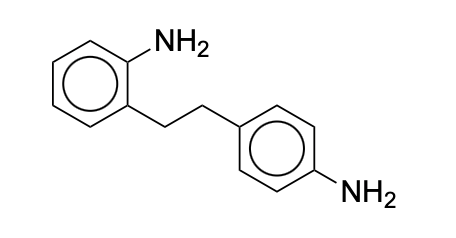}\\[1ex]
\includegraphics[width=5cm]{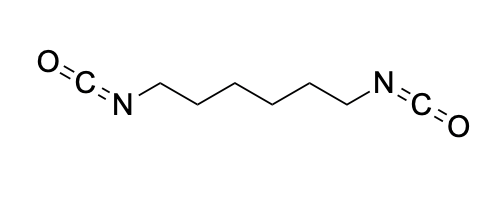}
} \\
\hline
\parbox{9.5cm}{
\small\textbf{SMILES:}\\
\texttt{[\textasteriskcentered]CCCCCCCCCCCSCCCCCCCCCCS([\textasteriskcentered])}\\[1.5ex]
\includegraphics[width=8.5cm]{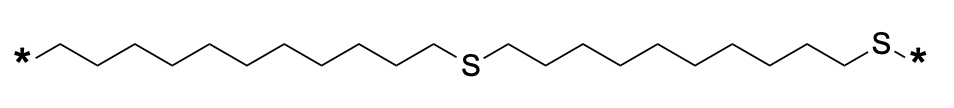}
} &
\parbox{6.5cm}{
\textbf{Addition}\\
Thiol $\Rightarrow$ Alkene.Thiol
} &
\parbox{6cm}{
\includegraphics[width=5cm]{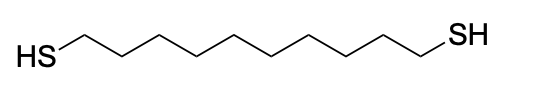}\\[1ex]
\includegraphics[width=5cm]{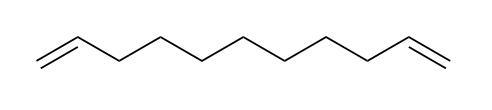}
} \\
\hline
\parbox{9.5cm}{
\small\textbf{SMILES:}\\
\texttt{[\textasteriskcentered]CCCNC1CC(=O)N(c2ccc(S(=O)(=O)c3ccc(N4C(=O)CC\\(NCCCN5CCN([\textasteriskcentered])CC5)C4=O)cc3)cc2)C1=O}\\[1.5ex]
\includegraphics[width=8.5cm]{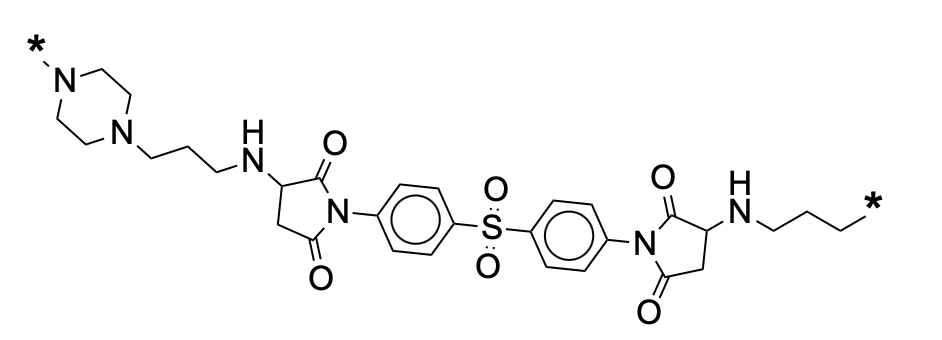}
} &
\parbox{6.5cm}{
\textbf{Addition}\\
Amine $\Rightarrow$ Alkene.Amine
} &
\parbox{6cm}{
\includegraphics[width=5cm]{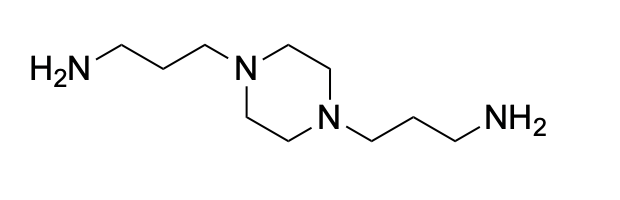}\\[1ex]
\includegraphics[width=5cm]{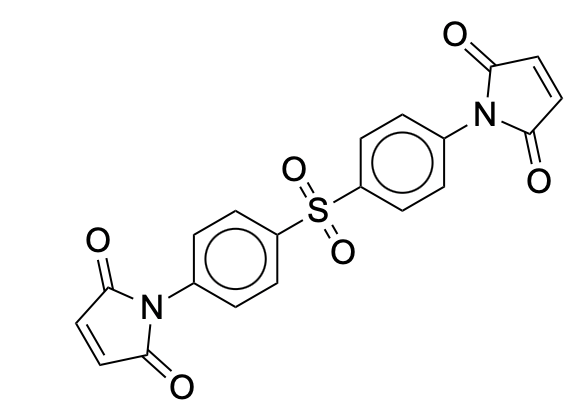}
} \\
\hline
\end{tabular}
\caption{Examples of the predicted polymerization class, reaction templates and the monomers using the polymer SMILES.}
\end{table}
\end{landscape}

\begin{landscape}
\begin{table}[h!]
\centering
\renewcommand{\arraystretch}{1.8}
\begin{tabular}{|P{9.5cm}|P{6.5cm}|P{6cm}|}
\hline
\textbf{Polymer } & \textbf{Language Template} & \textbf{Monomers} \\
\hline
\parbox{9.5cm}{
\small\textbf{SMILES:}\\
\texttt{[\textasteriskcentered][*]CC(C)S([\textasteriskcentered])(=O)=O}\\[1ex]
\includegraphics[width=3.5cm]{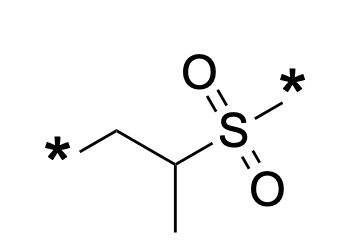}
} &
\parbox{6.5cm}{
\textbf{Addition}\\
Sulfone $\Rightarrow$ Alkane.Sulfone
} &
\parbox{6cm}{
\includegraphics[width=2cm]{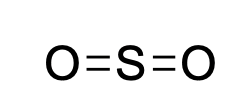}\\[1ex]
\includegraphics[width=2cm]{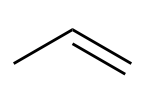}
} \\
\hline
\parbox{9.5cm}{
\small\textbf{SMILES:}\\
\texttt{[\textasteriskcentered]c1cccc(NC=CC(=O)c2ccc(C(=O)C=CNc3cccc(S([\textasteriskcentered])\\(=O)=O)c3)cc2)c1}\\[1.5ex]
\includegraphics[width=8.5cm]{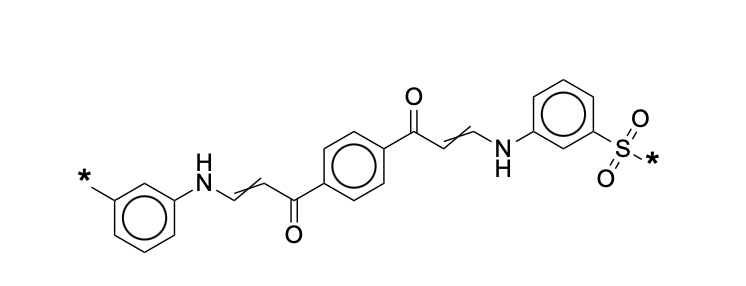}
} &
\parbox{6.5cm}{
\textbf{Addition}\\
Nitrile $\Rightarrow$ Alkyne.Amine
} &
\parbox{6cm}{
\includegraphics[width=5cm]{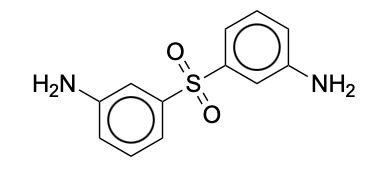}\\[1ex]
\includegraphics[width=5cm]{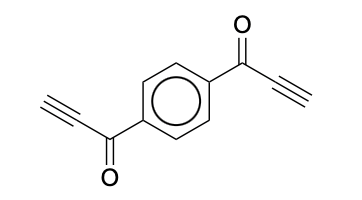}
} \\
\hline
\parbox{9.5cm}{
\small\textbf{SMILES:}\\
\texttt{[\textasteriskcentered]C=C[Si]([\textasteriskcentered])(c1ccccc1)c1ccccc1}\\[1.5ex]
\includegraphics[width=5.5cm]{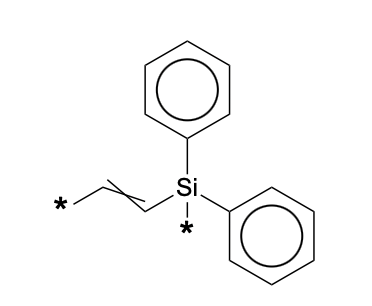}
} &
\parbox{6.5cm}{
\textbf{Addition}\\
Silane $\Rightarrow$ Alkyne.Silane
} &
\parbox{6cm}{
\includegraphics[width=3cm]{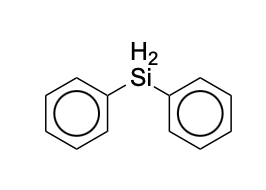}\\[1ex]
\includegraphics[width=3cm]{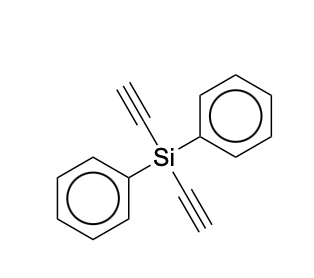}
} \\
\hline
\end{tabular}
\caption{Examples of the predicted polymerization class, reaction templates and the monomers using the polymer SMILES.}
\end{table}
\end{landscape}

\begin{landscape}
\begin{table}[h!]
\centering
\renewcommand{\arraystretch}{1.8}
\begin{tabular}{|P{9.5cm}|P{6.5cm}|P{6cm}|}
\hline
\textbf{Polymer } & \textbf{Language Template} & \textbf{Monomers} \\
\hline
\parbox{9.5cm}{
\small\textbf{SMILES:}\\
\texttt{[\textasteriskcentered][*]CC(C)S([\textasteriskcentered])(=O)=O}\\[1ex]
\includegraphics[width=5.5cm]{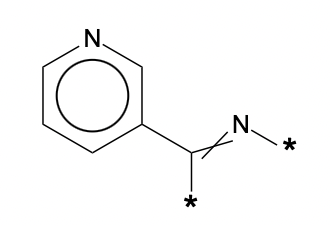}
} &
\parbox{6.5cm}{
\textbf{Addition}\\
Imine $\Rightarrow$ Imine.Imine
} &
\parbox{6cm}{
\includegraphics[width=3cm]{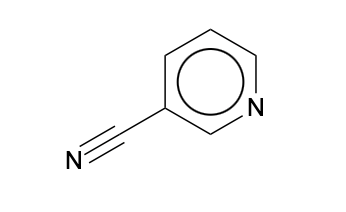}\\[1ex]
\includegraphics[width=3cm]{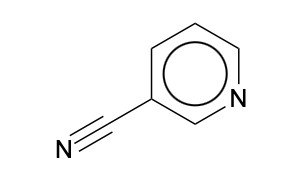}
} \\
\hline
\parbox{9.5cm}{
\small\textbf{SMILES:}\\
\texttt{[\textasteriskcentered]c1cccc(NC=CC(=O)c2ccc(C(=O)C=CNc3cccc(S([\textasteriskcentered])\\(=O)=O)c3)cc2)c1}\\[1.5ex]
\includegraphics[width=5.5cm]{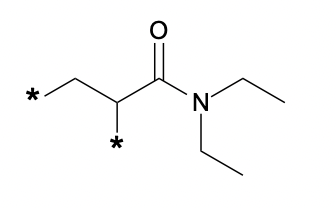}
} &
\parbox{6.5cm}{
\textbf{Addition}\\
Alkane $\Rightarrow$ Alkene.Alkene
} &
\parbox{6cm}{
\includegraphics[width=3cm]{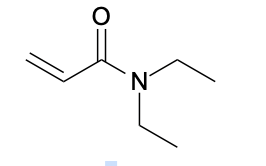}\\[1ex]
\includegraphics[width=3cm]{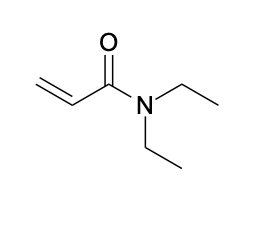}
} \\
\hline
\parbox{9.5cm}{
\small\textbf{SMILES:}\\
\texttt{[\textasteriskcentered]C=C[Si]([\textasteriskcentered])(c1ccccc1)c1ccccc1}\\[1.5ex]
\includegraphics[width=8.5cm]{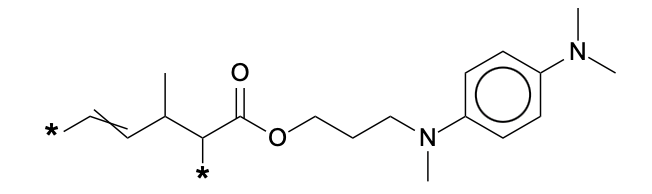}
} &
\parbox{6.5cm}{
\textbf{Addition}\\
Alkene $\Rightarrow$ Alkene.Alkene
} &
\parbox{6cm}{
\includegraphics[width=6cm]{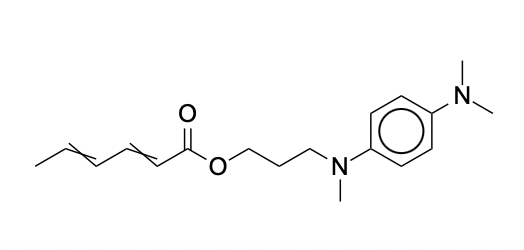}\\[1ex]
\includegraphics[width=6cm]{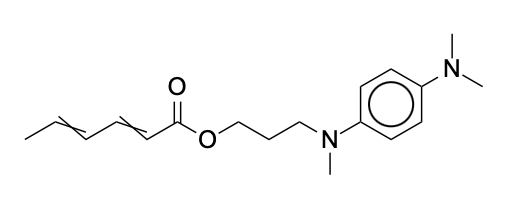}
} \\
\hline
\end{tabular}
\caption{Examples of the predicted polymerization class, reaction templates and the monomers using the polymer SMILES.}
\end{table}
\end{landscape}

\begin{landscape}
\begin{table}[h!]
\centering
\renewcommand{\arraystretch}{1.8}
\begin{tabular}{|P{9.5cm}|P{6.5cm}|P{6cm}|}
\hline
\textbf{Polymer } & \textbf{Language Template} & \textbf{Monomers} \\
\hline
\parbox{9.5cm}{
\small\textbf{SMILES:}\\
\texttt{[\textasteriskcentered][*]CC(C)S([\textasteriskcentered])(=O)=O}\\[1ex]
\includegraphics[width=5.5cm]{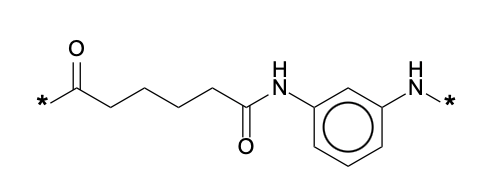}
} &
\parbox{6.5cm}{
\textbf{Addition}\\
Amide $\Rightarrow$ Carboxylic acid.Isocynate
} &
\parbox{6cm}{
\includegraphics[width=3cm]{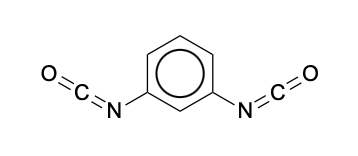}\\[1ex]
\includegraphics[width=3cm]{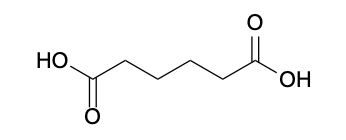}
} \\
\hline
\parbox{9.5cm}{
\small\textbf{SMILES:}\\
\texttt{[\textasteriskcentered]c1cccc(NC=CC(=O)c2ccc(C(=O)C=CNc3cccc(S([\textasteriskcentered])\\(=O)=O)c3)cc2)c1}\\[1.5ex]
\includegraphics[width=8.5cm]{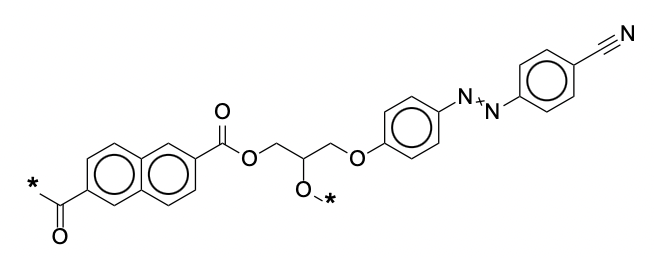}
} &
\parbox{6.5cm}{
\textbf{Addition}\\
Ester $\Rightarrow$ Alcohol.Carboxylic acid
} &
\parbox{6cm}{
\includegraphics[width=5cm]{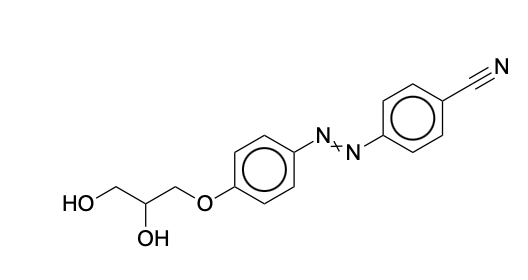}\\[1ex]
\includegraphics[width=5cm]{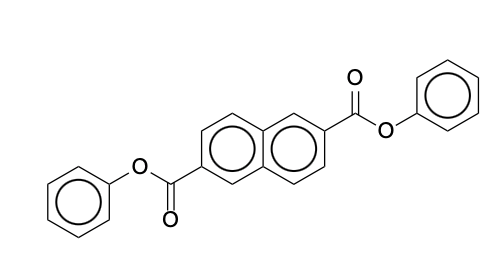}
} \\
\hline
\parbox{9.5cm}{
\small\textbf{SMILES:}\\
\texttt{[\textasteriskcentered]C=C[Si]([\textasteriskcentered])(c1ccccc1)c1ccccc1}\\[1.5ex]
\includegraphics[width=5.5cm]{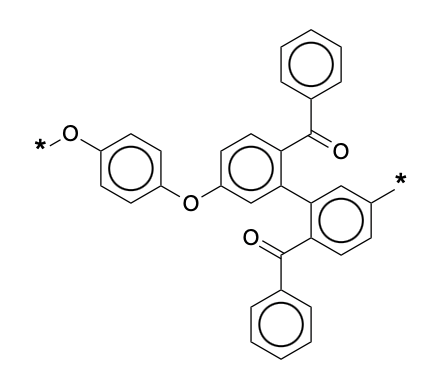}
} &
\parbox{6.5cm}{
\textbf{Addition}\\
Ether $\Rightarrow$ Alcohol.Halide
} &
\parbox{6cm}{
\includegraphics[width=3cm]{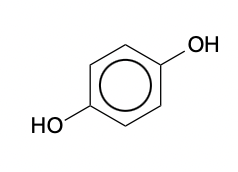}\\[1ex]
\includegraphics[width=3cm]{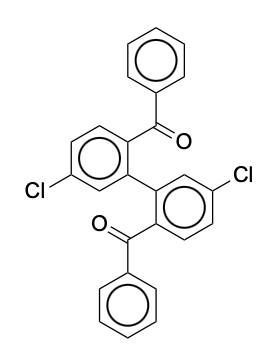}
} \\
\hline
\end{tabular}
\caption{Examples of the predicted polymerization class, reaction templates and the monomers using the polymer SMILES.}
\end{table}
\end{landscape}

\begin{landscape}
\begin{table}[h!]
\centering
\renewcommand{\arraystretch}{1.8}
\begin{tabular}{|P{9.5cm}|P{6.5cm}|P{6cm}|}
\hline
\textbf{Polymer } & \textbf{Language Template} & \textbf{Monomers} \\
\hline
\parbox{9.5cm}{
\small\textbf{SMILES:}\\
\texttt{[\textasteriskcentered][*]CC(C)S([\textasteriskcentered])(=O)=O}\\[1ex]
\includegraphics[width=5.5cm]{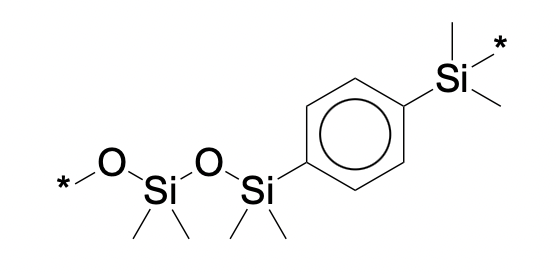}
} &
\parbox{6.5cm}{
\textbf{Condensation}\\
Alcohol $\Rightarrow$ Alcohol.Halide
} &
\parbox{6cm}{
\includegraphics[width=3cm]{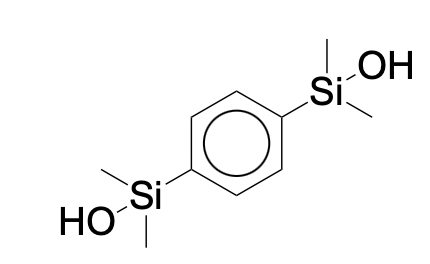}\\[1ex]
\includegraphics[width=3cm]{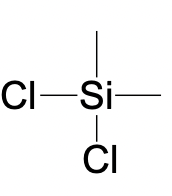}
} \\
\hline
\parbox{9.5cm}{
\small\textbf{SMILES:}\\
\texttt{[\textasteriskcentered]c1cccc(NC=CC(=O)c2ccc(C(=O)C=CNc3cccc(S([\textasteriskcentered])\\(=O)=O)c3)cc2)c1}\\[1.5ex]
\includegraphics[width=5cm]{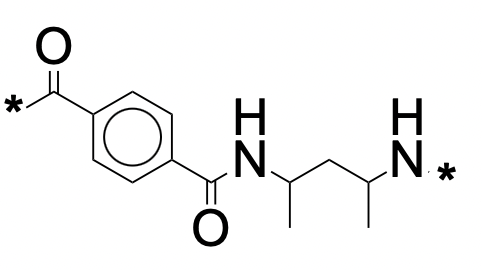}
} &
\parbox{6.5cm}{
\textbf{Condensation}\\
Amide $\Rightarrow$ Amine.Carboxylic acid
} &
\parbox{6cm}{
\includegraphics[width=3cm]{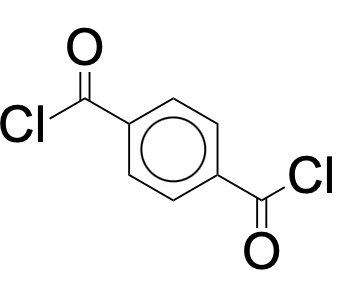}\\[1ex]
\includegraphics[width=3cm]{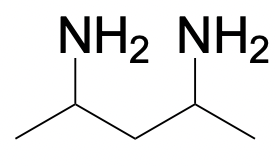}
} \\
\hline
\parbox{9.5cm}{
\small\textbf{SMILES:}\\
\texttt{[\textasteriskcentered]C=C[Si]([\textasteriskcentered])(c1ccccc1)c1ccccc1}\\[1.5ex]
\includegraphics[width=5.5cm]{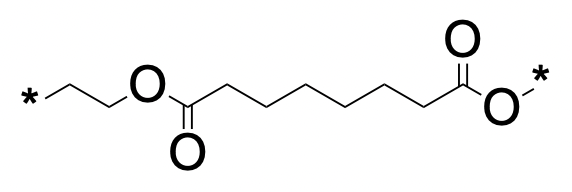}
} &
\parbox{6.5cm}{
\textbf{Condensation}\\
Ester $\Rightarrow$ Alcohol.Carboxylic acid
} &
\parbox{6cm}{
\includegraphics[width=3cm]{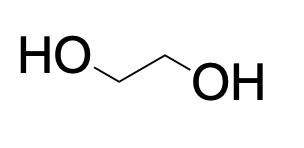}\\[1ex]
\includegraphics[width=3cm]{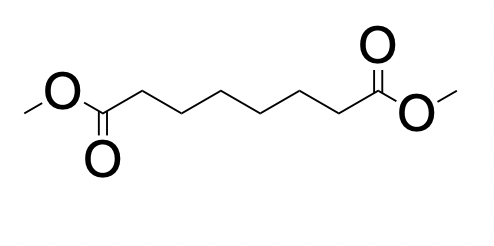}
} \\
\hline
\end{tabular}
\caption{Examples of the predicted polymerization class, reaction templates and the monomers using the polymer SMILES.}
\end{table}
\end{landscape}

\begin{landscape}
\begin{table}[h!]
\centering
\renewcommand{\arraystretch}{1.8}
\begin{tabular}{|P{9.5cm}|P{6.5cm}|P{6cm}|}
\hline
\textbf{Polymer } & \textbf{Language Template} & \textbf{Monomers} \\
\hline
\parbox{9.5cm}{
\small\textbf{SMILES:}\\
\texttt{[\textasteriskcentered][*]CC(C)S([\textasteriskcentered])(=O)=O}\\[1ex]
\includegraphics[width=5.5cm]{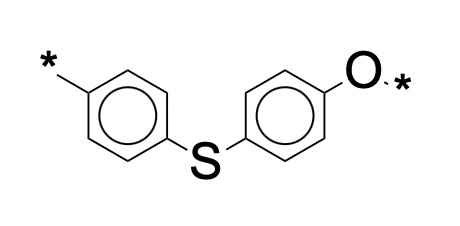}
} &
\parbox{6.5cm}{
\textbf{Condensation}\\
Thioether $\Rightarrow$ Thiol.Halide
} &
\parbox{6cm}{
\includegraphics[width=3cm]{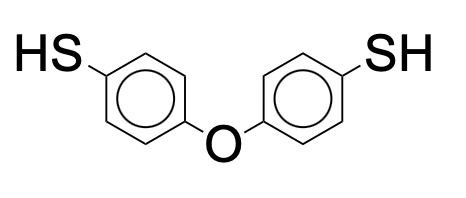}\\[1ex]
\includegraphics[width=3cm]{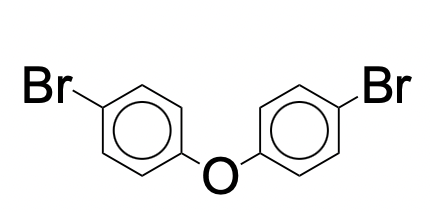}
} \\
\hline
\parbox{9.5cm}{
\small\textbf{SMILES:}\\
\texttt{[\textasteriskcentered]c1cccc(NC=CC(=O)c2ccc(C(=O)C=CNc3cccc(S([\textasteriskcentered])\\(=O)=O)c3)cc2)c1}\\[1.5ex]
\includegraphics[width=5cm]{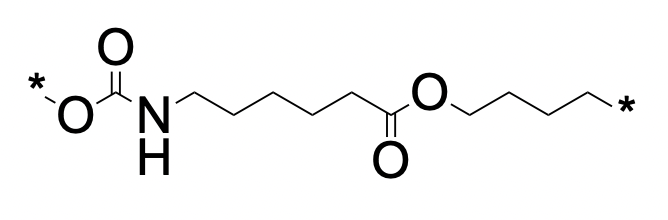}
} &
\parbox{6.5cm}{
\textbf{Condensation}\\
Urethane $\Rightarrow$ Alcohol.Isocynate
} &
\parbox{6cm}{
\includegraphics[width=3cm]{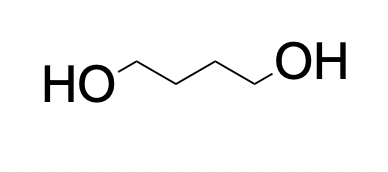}\\[1ex]
\includegraphics[width=3cm]{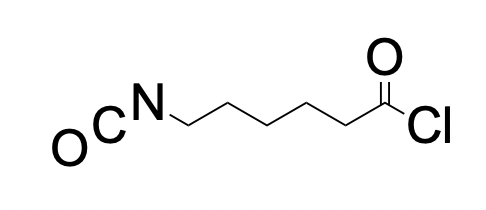}
} \\
\hline
\parbox{9.5cm}{
\small\textbf{SMILES:}\\
\texttt{[\textasteriskcentered]C=C[Si]([\textasteriskcentered])(c1ccccc1)c1ccccc1}\\[1.5ex]
\includegraphics[width=5.5cm]{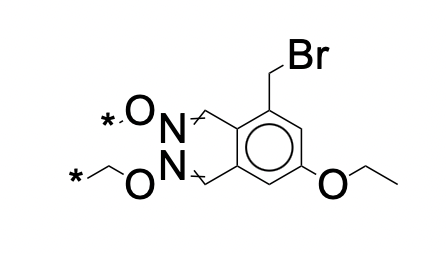}
} &
\parbox{6.5cm}{
\textbf{Condensation}\\
Oxime $\Rightarrow$ Hydroxylamine.Ketone(aldehyde)
} &
\parbox{6cm}{
\includegraphics[width=3cm]{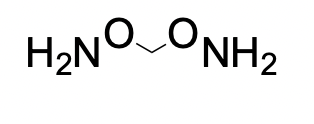}\\[1ex]
\includegraphics[width=3cm]{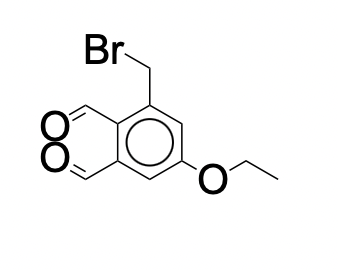}
} \\
\hline
\parbox{9.5cm}{
\small\textbf{SMILES:}\\
\texttt{[\textasteriskcentered]C=C[Si]([\textasteriskcentered])(c1ccccc1)c1ccccc1}\\[1.5ex]
\includegraphics[width=5.5cm]{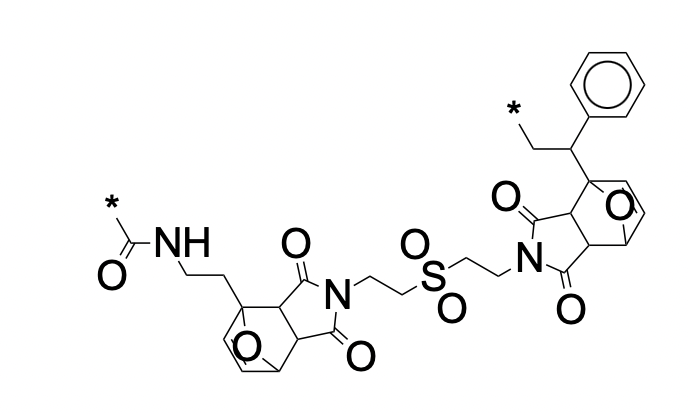}
} &
\parbox{6.5cm}{
\textbf{Condensation}\\
Furan maleimide $\Rightarrow$ Furan.Maleimide
} &
\parbox{6cm}{
\includegraphics[width=3cm]{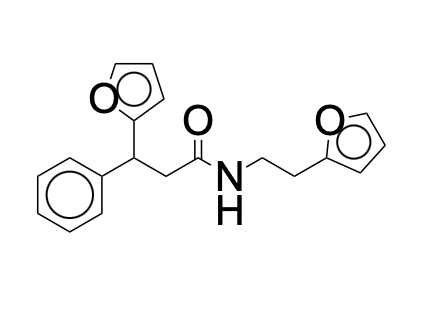}\\[1ex]
\includegraphics[width=3cm]{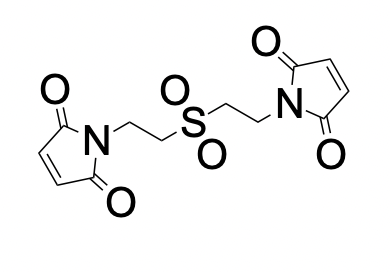}
} \\
\hline
\end{tabular}
\caption{Examples of the predicted polymerization class, reaction templates and the monomers using the polymer SMILES.}
\end{table}
\end{landscape}

\begin{landscape}
\begin{table}[h!]
\centering
\renewcommand{\arraystretch}{1.8}
\begin{tabular}{|P{9.5cm}|P{6.5cm}|P{6cm}|}
\hline
\textbf{Polymer } & \textbf{Language Template} & \textbf{Monomers} \\
\hline
\parbox{9.5cm}{
\small\textbf{SMILES:}\\
\texttt{[\textasteriskcentered][*]CC(C)S([\textasteriskcentered])(=O)=O}\\[1ex]
\includegraphics[width=5.5cm]{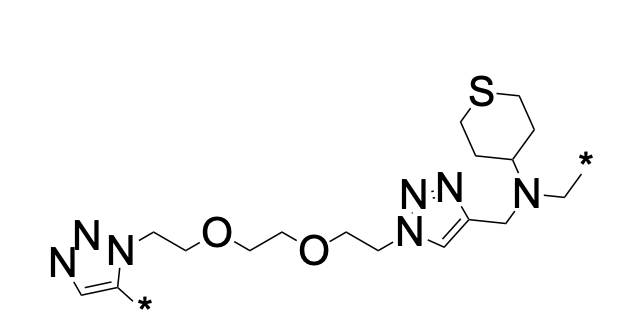}
} &
\parbox{6.5cm}{
\textbf{Condensation}\\
Triazole $\Rightarrow$ Alkyne.Azide
} &
\parbox{6cm}{
\includegraphics[width=3cm]{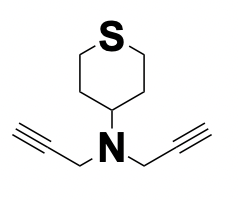}\\[1ex]
\includegraphics[width=3cm]{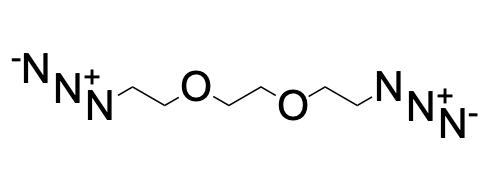}
} \\
\hline
\parbox{9.5cm}{
\small\textbf{SMILES:}\\
\texttt{[\textasteriskcentered]c1cccc(NC=CC(=O)c2ccc(C(=O)C=CNc3cccc(S([\textasteriskcentered])\\(=O)=O)c3)cc2)c1}\\[1.5ex]
\includegraphics[width=5cm]{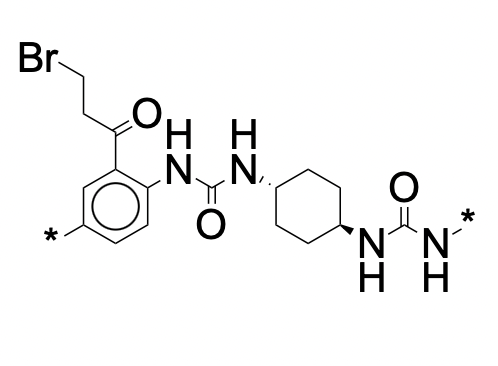}
} &
\parbox{6.5cm}{
\textbf{Condensation}\\
Urea $\Rightarrow$ Amine.Isocynate
} &
\parbox{6cm}{
\includegraphics[width=3cm]{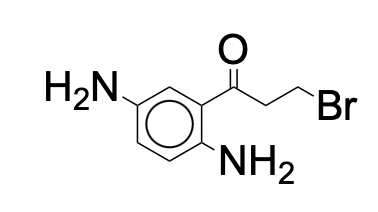}\\[1ex]
\includegraphics[width=3cm]{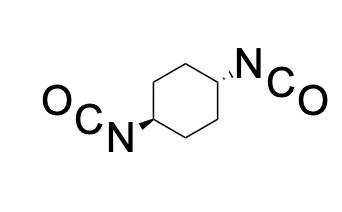}
} \\
\hline
\parbox{9.5cm}{
\small\textbf{SMILES:}\\
\texttt{[\textasteriskcentered]C=C[Si]([\textasteriskcentered])(c1ccccc1)c1ccccc1}\\[1.5ex]
\includegraphics[width=5.5cm]{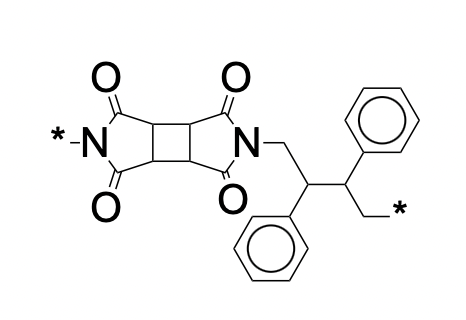}
} &
\parbox{6.5cm}{
\textbf{Condensation}\\
Imide $\Rightarrow$ Amine.Anhydride
} &
\parbox{6cm}{
\includegraphics[width=3cm]{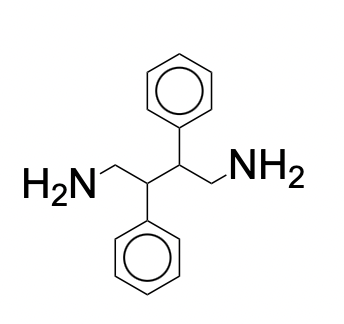}\\[1ex]
\includegraphics[width=3cm]{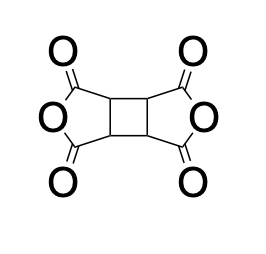}
} \\
\hline

\end{tabular}
\caption{Examples of the predicted polymerization class, reaction templates and the monomers using the polymer SMILES.}
\end{table}
\end{landscape}

\bibliography{achemso-demo}

\providecommand{\latin}[1]{#1}
\makeatletter
\providecommand{\doi}
  {\begingroup\let\do\@makeother\dospecials
  \catcode`\{=1 \catcode`\}=2 \doi@aux}
\providecommand{\doi@aux}[1]{\endgroup\texttt{#1}}
\makeatother
\providecommand*\mcitethebibliography{\thebibliography}
\csname @ifundefined\endcsname{endmcitethebibliography}  {\let\endmcitethebibliography\endthebibliography}{}
\begin{mcitethebibliography}{31}
\providecommand*\natexlab[1]{#1}
\providecommand*\mciteSetBstSublistMode[1]{}
\providecommand*\mciteSetBstMaxWidthForm[2]{}
\providecommand*\mciteBstWouldAddEndPuncttrue
  {\def\EndOfBibitem{\unskip.}}
\providecommand*\mciteBstWouldAddEndPunctfalse
  {\let\EndOfBibitem\relax}
\providecommand*\mciteSetBstMidEndSepPunct[3]{}
\providecommand*\mciteSetBstSublistLabelBeginEnd[3]{}
\providecommand*\EndOfBibitem{}
\mciteSetBstSublistMode{f}
\mciteSetBstMaxWidthForm{subitem}{(\alph{mcitesubitemcount})}
\mciteSetBstSublistLabelBeginEnd
  {\mcitemaxwidthsubitemform\space}
  {\relax}
  {\relax}

\bibitem[Miller-Chou and Koenig(2003)Miller-Chou, and Koenig]{miller2003review}
Miller-Chou,~B.~A.; Koenig,~J.~L. A review of polymer dissolution. \emph{Progress in polymer science} \textbf{2003}, \emph{28}, 1223--1270\relax
\mciteBstWouldAddEndPuncttrue
\mciteSetBstMidEndSepPunct{\mcitedefaultmidpunct}
{\mcitedefaultendpunct}{\mcitedefaultseppunct}\relax
\EndOfBibitem
\bibitem[Acikgoz \latin{et~al.}(2011)Acikgoz, Hempenius, Huskens, and Vancso]{acikgoz2011polymers}
Acikgoz,~C.; Hempenius,~M.~A.; Huskens,~J.; Vancso,~G.~J. Polymers in conventional and alternative lithography for the fabrication of nanostructures. \emph{European Polymer Journal} \textbf{2011}, \emph{47}, 2033--2052\relax
\mciteBstWouldAddEndPuncttrue
\mciteSetBstMidEndSepPunct{\mcitedefaultmidpunct}
{\mcitedefaultendpunct}{\mcitedefaultseppunct}\relax
\EndOfBibitem
\bibitem[Dong \latin{et~al.}(2021)Dong, Lu, Harris, and Escobar]{dong2021polymers}
Dong,~X.; Lu,~D.; Harris,~T.~A.; Escobar,~I.~C. Polymers and solvents used in membrane fabrication: a review focusing on sustainable membrane development. \emph{Membranes} \textbf{2021}, \emph{11}, 309\relax
\mciteBstWouldAddEndPuncttrue
\mciteSetBstMidEndSepPunct{\mcitedefaultmidpunct}
{\mcitedefaultendpunct}{\mcitedefaultseppunct}\relax
\EndOfBibitem
\bibitem[Choi \latin{et~al.}(2022)Choi, Woo, Choi, and Jin]{choi2022effects}
Choi,~M.-J.; Woo,~M.~R.; Choi,~H.-G.; Jin,~S.~G. Effects of polymers on the drug solubility and dissolution enhancement of poorly water-soluble rivaroxaban. \emph{International Journal of Molecular Sciences} \textbf{2022}, \emph{23}, 9491\relax
\mciteBstWouldAddEndPuncttrue
\mciteSetBstMidEndSepPunct{\mcitedefaultmidpunct}
{\mcitedefaultendpunct}{\mcitedefaultseppunct}\relax
\EndOfBibitem
\bibitem[Medarevi{\'c} \latin{et~al.}(2019)Medarevi{\'c}, Djuri{\v{s}}, Barmpalexis, Kachrimanis, and Ibri{\'c}]{medarevic2019analytical}
Medarevi{\'c},~D.; Djuri{\v{s}},~J.; Barmpalexis,~P.; Kachrimanis,~K.; Ibri{\'c},~S. Analytical and computational methods for the estimation of drug-polymer solubility and miscibility in solid dispersions development. \emph{Pharmaceutics} \textbf{2019}, \emph{11}, 372\relax
\mciteBstWouldAddEndPuncttrue
\mciteSetBstMidEndSepPunct{\mcitedefaultmidpunct}
{\mcitedefaultendpunct}{\mcitedefaultseppunct}\relax
\EndOfBibitem
\bibitem[Kadajji and Betageri(2011)Kadajji, and Betageri]{kadajji2011water}
Kadajji,~V.~G.; Betageri,~G.~V. Water soluble polymers for pharmaceutical applications. \emph{Polymers} \textbf{2011}, \emph{3}, 1972--2009\relax
\mciteBstWouldAddEndPuncttrue
\mciteSetBstMidEndSepPunct{\mcitedefaultmidpunct}
{\mcitedefaultendpunct}{\mcitedefaultseppunct}\relax
\EndOfBibitem
\bibitem[Tsampanakis and Orbaek~White(2021)Tsampanakis, and Orbaek~White]{tsampanakis2021mechanics}
Tsampanakis,~I.; Orbaek~White,~A. The mechanics of forming ideal polymer--solvent combinations for open-loop chemical recycling of solvents and plastics. \emph{Polymers} \textbf{2021}, \emph{14}, 112\relax
\mciteBstWouldAddEndPuncttrue
\mciteSetBstMidEndSepPunct{\mcitedefaultmidpunct}
{\mcitedefaultendpunct}{\mcitedefaultseppunct}\relax
\EndOfBibitem
\bibitem[Rivas \latin{et~al.}(2018)Rivas, Urbano, and S{\'a}nchez]{rivas2018water}
Rivas,~B.~L.; Urbano,~B.~F.; S{\'a}nchez,~J. Water-soluble and insoluble polymers, nanoparticles, nanocomposites and hybrids with ability to remove hazardous inorganic pollutants in water. \emph{Frontiers in chemistry} \textbf{2018}, \emph{6}, 320\relax
\mciteBstWouldAddEndPuncttrue
\mciteSetBstMidEndSepPunct{\mcitedefaultmidpunct}
{\mcitedefaultendpunct}{\mcitedefaultseppunct}\relax
\EndOfBibitem
\bibitem[Duis \latin{et~al.}(2021)Duis, Junker, and Coors]{duis2021environmental}
Duis,~K.; Junker,~T.; Coors,~A. Environmental fate and effects of water-soluble synthetic organic polymers used in cosmetic products. \emph{Environmental Sciences Europe} \textbf{2021}, \emph{33}, 21\relax
\mciteBstWouldAddEndPuncttrue
\mciteSetBstMidEndSepPunct{\mcitedefaultmidpunct}
{\mcitedefaultendpunct}{\mcitedefaultseppunct}\relax
\EndOfBibitem
\bibitem[Barton(1975)]{barton1975solubility}
Barton,~A.~F. Solubility parameters. \emph{Chemical Reviews} \textbf{1975}, \emph{75}, 731--753\relax
\mciteBstWouldAddEndPuncttrue
\mciteSetBstMidEndSepPunct{\mcitedefaultmidpunct}
{\mcitedefaultendpunct}{\mcitedefaultseppunct}\relax
\EndOfBibitem
\bibitem[Thakral and Thakral(2013)Thakral, and Thakral]{thakral2013prediction}
Thakral,~S.; Thakral,~N.~K. Prediction of drug--polymer miscibility through the use of solubility parameter based Flory--Huggins interaction parameter and the experimental validation: PEG as model polymer. \emph{Journal of pharmaceutical sciences} \textbf{2013}, \emph{102}, 2254--2263\relax
\mciteBstWouldAddEndPuncttrue
\mciteSetBstMidEndSepPunct{\mcitedefaultmidpunct}
{\mcitedefaultendpunct}{\mcitedefaultseppunct}\relax
\EndOfBibitem
\bibitem[Sanchez-Lengeling \latin{et~al.}(2019)Sanchez-Lengeling, Roch, Perea, Langner, Brabec, and Aspuru-Guzik]{sanchez2019bayesian}
Sanchez-Lengeling,~B.; Roch,~L.~M.; Perea,~J.~D.; Langner,~S.; Brabec,~C.~J.; Aspuru-Guzik,~A. A Bayesian approach to predict solubility parameters. \emph{Advanced Theory and Simulations} \textbf{2019}, \emph{2}, 1800069\relax
\mciteBstWouldAddEndPuncttrue
\mciteSetBstMidEndSepPunct{\mcitedefaultmidpunct}
{\mcitedefaultendpunct}{\mcitedefaultseppunct}\relax
\EndOfBibitem
\bibitem[Kurotani \latin{et~al.}(2021)Kurotani, Kakiuchi, and Kikuchi]{kurotani2021solubility}
Kurotani,~A.; Kakiuchi,~T.; Kikuchi,~J. Solubility prediction from molecular properties and analytical data using an in-phase deep neural network (Ip-DNN). \emph{ACS omega} \textbf{2021}, \emph{6}, 14278--14287\relax
\mciteBstWouldAddEndPuncttrue
\mciteSetBstMidEndSepPunct{\mcitedefaultmidpunct}
{\mcitedefaultendpunct}{\mcitedefaultseppunct}\relax
\EndOfBibitem
\bibitem[Chi \latin{et~al.}(2021)Chi, Gargouri, Schrader, Damak, Ma{\^a}lej, and Sierka]{chi2021atomistic}
Chi,~M.; Gargouri,~R.; Schrader,~T.; Damak,~K.; Ma{\^a}lej,~R.; Sierka,~M. Atomistic descriptors for machine learning models of solubility parameters for small molecules and polymers. \emph{Polymers} \textbf{2021}, \emph{14}, 26\relax
\mciteBstWouldAddEndPuncttrue
\mciteSetBstMidEndSepPunct{\mcitedefaultmidpunct}
{\mcitedefaultendpunct}{\mcitedefaultseppunct}\relax
\EndOfBibitem
\bibitem[Liu \latin{et~al.}(2022)Liu, Liu, Ding, and Li]{liu2022machine}
Liu,~T.-L.; Liu,~L.-Y.; Ding,~F.; Li,~Y.-Q. A machine learning study of polymer-solvent interactions. \emph{Chinese Journal of Polymer Science} \textbf{2022}, \emph{40}, 834--842\relax
\mciteBstWouldAddEndPuncttrue
\mciteSetBstMidEndSepPunct{\mcitedefaultmidpunct}
{\mcitedefaultendpunct}{\mcitedefaultseppunct}\relax
\EndOfBibitem
\bibitem[Chandrasekaran \latin{et~al.}(2020)Chandrasekaran, Kim, Venkatram, and Ramprasad]{chandrasekaran2020deep}
Chandrasekaran,~A.; Kim,~C.; Venkatram,~S.; Ramprasad,~R. A deep learning solvent-selection paradigm powered by a massive solvent/nonsolvent database for polymers. \emph{Macromolecules} \textbf{2020}, \emph{53}, 4764--4769\relax
\mciteBstWouldAddEndPuncttrue
\mciteSetBstMidEndSepPunct{\mcitedefaultmidpunct}
{\mcitedefaultendpunct}{\mcitedefaultseppunct}\relax
\EndOfBibitem
\bibitem[Chen \latin{et~al.}(2021)Chen, Pilania, Batra, Huan, Kim, Kuenneth, and Ramprasad]{chen2021polymer}
Chen,~L.; Pilania,~G.; Batra,~R.; Huan,~T.~D.; Kim,~C.; Kuenneth,~C.; Ramprasad,~R. Polymer informatics: Current status and critical next steps. \emph{Materials Science and Engineering: R: Reports} \textbf{2021}, \emph{144}, 100595\relax
\mciteBstWouldAddEndPuncttrue
\mciteSetBstMidEndSepPunct{\mcitedefaultmidpunct}
{\mcitedefaultendpunct}{\mcitedefaultseppunct}\relax
\EndOfBibitem
\bibitem[Kern \latin{et~al.}(2022)Kern, Venkatram, Banerjee, Brettmann, and Ramprasad]{kern2022solvent}
Kern,~J.; Venkatram,~S.; Banerjee,~M.; Brettmann,~B.; Ramprasad,~R. Solvent selection for polymers enabled by generalized chemical fingerprinting and machine learning. \emph{Physical Chemistry Chemical Physics} \textbf{2022}, \emph{24}, 26547--26555\relax
\mciteBstWouldAddEndPuncttrue
\mciteSetBstMidEndSepPunct{\mcitedefaultmidpunct}
{\mcitedefaultendpunct}{\mcitedefaultseppunct}\relax
\EndOfBibitem
\bibitem[Tran \latin{et~al.}(2024)Tran, Gurnani, Kim, Pilania, Kwon, Lively, and Ramprasad]{tran2024design}
Tran,~H.; Gurnani,~R.; Kim,~C.; Pilania,~G.; Kwon,~H.-K.; Lively,~R.~P.; Ramprasad,~R. Design of functional and sustainable polymers assisted by artificial intelligence. \emph{Nature Reviews Materials} \textbf{2024}, 1--21\relax
\mciteBstWouldAddEndPuncttrue
\mciteSetBstMidEndSepPunct{\mcitedefaultmidpunct}
{\mcitedefaultendpunct}{\mcitedefaultseppunct}\relax
\EndOfBibitem
\bibitem[Zhang \latin{et~al.}(2023)Zhang, Zhou, Ming, and Sun]{zhang2023gpt}
Zhang,~X.; Zhou,~Z.; Ming,~C.; Sun,~Y.-Y. GPT-Assisted Learning of Structure--Property Relationships by Graph Neural Networks: Application to Rare-Earth-Doped Phosphors. \emph{The Journal of Physical Chemistry Letters} \textbf{2023}, \emph{14}, 11342--11349\relax
\mciteBstWouldAddEndPuncttrue
\mciteSetBstMidEndSepPunct{\mcitedefaultmidpunct}
{\mcitedefaultendpunct}{\mcitedefaultseppunct}\relax
\EndOfBibitem
\bibitem[Jablonka \latin{et~al.}(2023)Jablonka, Schwaller, Ortega-Guerrero, and Smit]{jablonka2023gpt}
Jablonka,~K.~M.; Schwaller,~P.; Ortega-Guerrero,~A.; Smit,~B. Is GPT-3 all you need for low-data discovery in chemistry? \textbf{2023}, \relax
\mciteBstWouldAddEndPunctfalse
\mciteSetBstMidEndSepPunct{\mcitedefaultmidpunct}
{}{\mcitedefaultseppunct}\relax
\EndOfBibitem
\bibitem[Xie \latin{et~al.}(2023)Xie, Wan, Huang, Zhou, Liu, Linghu, Wang, Kit, Grazian, Zhang, \latin{et~al.} others]{xie2023large}
Xie,~T.; Wan,~Y.; Huang,~W.; Zhou,~Y.; Liu,~Y.; Linghu,~Q.; Wang,~S.; Kit,~C.; Grazian,~C.; Zhang,~W.; others Large language models as master key: unlocking the secrets of materials science with GPT. \emph{arXiv preprint arXiv:2304.02213} \textbf{2023}, \relax
\mciteBstWouldAddEndPunctfalse
\mciteSetBstMidEndSepPunct{\mcitedefaultmidpunct}
{}{\mcitedefaultseppunct}\relax
\EndOfBibitem
\bibitem[Dai \latin{et~al.}(2025)Dai, Zhang, Wei, Lin, Dai, Peng, Song, Tang, Li, Liu, \latin{et~al.} others]{dai2025gpt}
Dai,~D.; Zhang,~G.; Wei,~X.; Lin,~Y.; Dai,~M.; Peng,~J.; Song,~N.; Tang,~Z.; Li,~S.; Liu,~J.; others A GPT-assisted iterative method for extracting domain knowledge from a large volume of literature of electromagnetic wave absorbing materials with limited manually annotated data. \emph{Computational Materials Science} \textbf{2025}, \emph{246}, 113431\relax
\mciteBstWouldAddEndPuncttrue
\mciteSetBstMidEndSepPunct{\mcitedefaultmidpunct}
{\mcitedefaultendpunct}{\mcitedefaultseppunct}\relax
\EndOfBibitem
\bibitem[Xie \latin{et~al.}(2024)Xie, Evangelopoulos, Omar, Troisi, Cooper, and Chen]{xie2024fine}
Xie,~Z.; Evangelopoulos,~X.; Omar,~{\"O}.~H.; Troisi,~A.; Cooper,~A.~I.; Chen,~L. Fine-tuning GPT-3 for machine learning electronic and functional properties of organic molecules. \emph{Chemical science} \textbf{2024}, \emph{15}, 500--510\relax
\mciteBstWouldAddEndPuncttrue
\mciteSetBstMidEndSepPunct{\mcitedefaultmidpunct}
{\mcitedefaultendpunct}{\mcitedefaultseppunct}\relax
\EndOfBibitem
\bibitem[Jablonka \latin{et~al.}(2024)Jablonka, Schwaller, Ortega-Guerrero, and Smit]{jablonka2024leveraging}
Jablonka,~K.~M.; Schwaller,~P.; Ortega-Guerrero,~A.; Smit,~B. Leveraging large language models for predictive chemistry. \emph{Nature Machine Intelligence} \textbf{2024}, \emph{6}, 161--169\relax
\mciteBstWouldAddEndPuncttrue
\mciteSetBstMidEndSepPunct{\mcitedefaultmidpunct}
{\mcitedefaultendpunct}{\mcitedefaultseppunct}\relax
\EndOfBibitem
\bibitem[Mishra \latin{et~al.}(2024)Mishra, Singh, Ahlawat, Zaki, Bihani, Grover, Mishra, Miret, Krishnan, \latin{et~al.} others]{mishra2024foundational}
Mishra,~V.; Singh,~S.; Ahlawat,~D.; Zaki,~M.; Bihani,~V.; Grover,~H.~S.; Mishra,~B.; Miret,~S.; Krishnan,~N.; others Foundational Large Language Models for Materials Research. \emph{arXiv preprint arXiv:2412.09560} \textbf{2024}, \relax
\mciteBstWouldAddEndPunctfalse
\mciteSetBstMidEndSepPunct{\mcitedefaultmidpunct}
{}{\mcitedefaultseppunct}\relax
\EndOfBibitem
\bibitem[Ock \latin{et~al.}(2023)Ock, Guntuboina, and Barati~Farimani]{ock2023catalyst}
Ock,~J.; Guntuboina,~C.; Barati~Farimani,~A. Catalyst energy prediction with catberta: Unveiling feature exploration strategies through large language models. \emph{ACS Catalysis} \textbf{2023}, \emph{13}, 16032--16044\relax
\mciteBstWouldAddEndPuncttrue
\mciteSetBstMidEndSepPunct{\mcitedefaultmidpunct}
{\mcitedefaultendpunct}{\mcitedefaultseppunct}\relax
\EndOfBibitem
\bibitem[Ramos \latin{et~al.}(2024)Ramos, Collison, and White]{ramos2024review}
Ramos,~M.~C.; Collison,~C.; White,~A.~D. A review of large language models and autonomous agents in chemistry. \emph{Chemical Science} \textbf{2024}, \relax
\mciteBstWouldAddEndPunctfalse
\mciteSetBstMidEndSepPunct{\mcitedefaultmidpunct}
{}{\mcitedefaultseppunct}\relax
\EndOfBibitem
\bibitem[Lee and Hsiang(2020)Lee, and Hsiang]{lee2020patent}
Lee,~J.-S.; Hsiang,~J. Patent claim generation by fine-tuning OpenAI GPT-2. \emph{World Patent Information} \textbf{2020}, \emph{62}, 101983\relax
\mciteBstWouldAddEndPuncttrue
\mciteSetBstMidEndSepPunct{\mcitedefaultmidpunct}
{\mcitedefaultendpunct}{\mcitedefaultseppunct}\relax
\EndOfBibitem
\bibitem[Brown(2020)]{brown2020language}
Brown,~T.~B. Language models are few-shot learners. \emph{arXiv preprint arXiv:2005.14165} \textbf{2020}, \relax
\mciteBstWouldAddEndPunctfalse
\mciteSetBstMidEndSepPunct{\mcitedefaultmidpunct}
{}{\mcitedefaultseppunct}\relax
\EndOfBibitem
\end{mcitethebibliography}
\end{document}